\documentclass[usenatbib,fleqn]{mn2e} 

\usepackage{amsmath}
\usepackage{booktabs}
\usepackage{color}
\usepackage{graphicx}
\usepackage{multicol}
\usepackage{rotating}
\usepackage{caption}
\usepackage{aas_macros}
\usepackage{tabularx}
\usepackage{tabu}
\usepackage{siunitx}

\newcommand{\xbf}{\begin{figure}}
\newcommand{\xef}{\end{figure}}

\newcommand{\be}{\begin{equation}}
\newcommand{\ee}{\end{equation}}

\newcommand{\citedir}{\citet}
\newcommand{\citeind}{\citep}

\newcommand{\bi}[1]{\textit{#1}}
\newcommand{\riz}{\bi{r},\bi{i},\bi{z}}

\usepackage{draftwatermark}

\newcommand{\euclidreq}{3 \times 10^{-4}}

\title{Implications of a wavelength dependent PSF for weak lensing measurements.}
\author[Martin Eriksen \& Henk Hoekstra]
{\parbox{\textwidth}{Martin Eriksen \& Henk Hoekstra}
\vspace{0.4cm}\\
Leiden Observatory, Leiden University, PO Box 9513, NL-2300 RA Leiden, Netherlands \\
}

\newcommand{\xfigure}[1]{
\begin{center}
\includegraphics[width=0.5\textwidth]{#1}
\end{center}
}

\newcommand{\zfigure}[1]{
\begin{center}
\includegraphics[width=0.5\textwidth]{#1}
\end{center}
}

\newcommand{\yfigure}[1]{
\begin{center}
\includegraphics[width=1.0\textwidth]{#1} 
\end{center}
} 

\newcommand{\rpsf}{R^2_{\text{PSF}}}

\begin{document}
\maketitle
\begin{abstract}
The convolution of galaxy images by the point-spread function (PSF) is the dominant source of bias for weak gravitational lensing studies, and an accurate estimate of the PSF is required to obtain unbiased shape measurements. The PSF estimate for a galaxy depends on its spectral energy distribution (SED), because the instrumental PSF is generally a function of the wavelength. In this paper we explore various approaches to determine the resulting `effective' PSF using broad-band data. Considering the {\it Euclid} mission as a reference, we find that standard SED template fitting methods result in biases that depend on source redshift, although this may be remedied if the algorithms can be optimised for this purpose. 
Using a machine-learning algorithm we show that, at least in principle,  the required accuracy can be achieved with the current survey parameters. It is also possible to account for the correlations between photometric redshift and PSF estimates that arise from the use of the same photometry. We explore  the impact of errors in photometric calibration, errors in the assumed wavelength dependence of the PSF model and limitations of the adopted template libraries. Our results indicate that the required accuracy for {\it Euclid} can be achieved using the data that are planned to determine photometric redshifts.
 
\end{abstract}


\section{Introduction}
The measurement of the distance-redshift relation using distant type
Ia supernovae led to the remarkable discovery that the expansion of
the Universe is accelerating \citeind{sn1,sn2}. Since then, this
finding has been confirmed by a wide range of observations, but there
is still no consensus on the underlying theory: options range from a
cosmological constant to a change of fundamental
physics.  To restrict the range of explanations, significant
observational progress is required, and to this end a wide variety of
observational probes and facilities are being studied and employed
\citeind{weinberg_probes}.

Of particular interest is weak gravitational lensing
\citeind{HJ08,Kilbinger15}: the statistics of the coherent distortions of
the images of distant galaxies by intervening structures can be
related to the underlying cosmological model. Measuring this lensing
signal as a function of source redshift can in principle lead to some
of the tightest constraints on cosmological parameters. The typical
change in galaxy shape is tiny compared to its intrinsic ellipticity,
and a precise measurement involves averaging over large samples of
galaxies. Moreover, gravitational lensing is not the only phenomenon
that can lead to observed correlations in the galaxy shapes: tidal
effects during structure formation may lead to intrinsic alignments,
which complicate the interpretation of the measurements
\citep[e.g.][]{Joachimi15,Kirk15}.

Perhaps the biggest challenge is that a range of instrumental effects
can overwhelm the lensing signal, unless carefully corrected for.  Of
these, the convolution of the galaxy images by the point spread
function (PSF) is typically dominant, but other effects may contribute
as well \citeind{massey13,cropper}. Hence, much effort has focussed on
an accurate correction for the PSF, which circularises the images, but
can also introduce alignments if it is anisotropic.  Despite these
technical difficulties, the lensing signal by large-scale structure,
commonly referred to as `cosmic shear', has now been robustly measured
\citep[see e.g.][for recent results]{heymans2012,Becker16, Hildebrandt17}.

The next generation of lensing surveys will cover much larger areas of
sky and aim to measure shapes of billions of galaxies. The Large
Synoptic Survey Telescope\footnote{http://www.lsst.org/}
\citep[LSST;][]{lsst1} will survey the sky repeatedly from the ground,
whereas {\it Euclid}\footnote{http://www.euclid-ec.org/}
\citep{euclidrb}, and the Wide-Field Infrared Survey
Telescope\footnote{http://wfirst.gsfc.nasa.gov/} \citep[WFIRST;][]{wfirst} will observe from space to avoid the blurring of
the images by the atmosphere. The dramatic reduction in statistical
uncertainties afforded by these new surveys needs to be matched by a
reduction in the level of residual systematics. Consequently, even in
diffraction-limited space-based observations, the PSF cannot be
ignored \citep{cropper}.

The PSF varies spatially due to misalignments of optical elements,
which also typically vary with time due to changes in thermal conditions
and, in the case of ground-based telescopes, due to changing
gravitational loads.  This can be modelled using the observations of
stars in the field-of-view. A complication is that the PSF generally
depends on wavelength; this effect is stronger for diffraction-limited
optics, but atmospheric differential chromatic refraction and the
turbulence in the atmosphere also depend on wavelength
\citep{meyers2015}. Hence, the observed PSFs depend on the spectral
energy distribution (SED) of the stars. Fortunately the SEDs of stars
are well-studied and relatively smooth, such that with limited
broad-band colour information the wavelength dependence can also be
included in the PSF model.

Each galaxy, however, is convolved by a PSF that depends on its SED in
the observed frame, the `effective' PSF. An incorrect estimate of this
PSF will lead to biases in the galaxy shape estimates and consequently
in the cosmological parameters. Hence it is not only important that
the wavelength dependent model for the PSF is accurate, but also that
the galaxy SED can be inferred sufficiently well. In this paper we
focus on the spatially averaged, or global, SED of the galaxy, but we
note that spatial variations lead to additional complications
\citep{voigt,semboloni_colorgrad}, which we do not consider
here. Examining the impact of the wavelength dependence is
particularly relevant for {\it Euclid}, because the PSF is not only
diffraction limited, but the shape measurements are based on optical
data obtained using an especially broad passband \citep[5500-9200\AA;][]{euclidrb} 
to maximise the number of galaxies for which shapes can be measured.

To study the expansion history and growth of structure, lensing
surveys measure the cosmic shear signal as a function of source
redshift. Measuring spectroscopic redshifts for such large numbers of
faint galaxies is too costly, but fortunately photometric redshifts
are adequate. These are obtained by complementing the shape
measurements with photometry in multiple filters, which can also
provide information on the observed SEDs. Whether such data are
adequate for the determination of the effective PSF for galaxies was
first studied by \citedir{cypriano} in the context of {\it Euclid}.

\cite{cypriano} examined two approaches to account for the wavelength
dependent PSF. First, they explored whether stars with similar colours
as the galaxies could be used. In general one does not expect the SEDs
of stars to match those of galaxies well over the broad redshift range
covered by {\it Euclid}. Nonetheless, this approach performed
reasonably well, albeit with significant biases for high redshift
galaxies. \cite{cypriano} obtained better results by training a neural
network on simulated SEDs and combining this with a model for the
wavelength dependence of the PSF. This allowed them to to predict the
PSF size as a function of the observed galaxy colours. Similarly,
\citedir{meyers2015} explored how machine learning techniques can be
used to account for atmospheric chromatic effects in ground-based
data.

In this paper we revisit the problem studied by \citet{cypriano} and
\citet{meyers2015}, with a particular focus on what data are required
to meet the requirements for {\it Euclid}.  This paper examines the
performance of the various approaches to estimate the effective PSF
size, using a more up-to-date formulation of requirements, as
presented in \cite{massey13}.  The detailed break down of various
sources of bias presented in \cite{cropper} indicates that the actual
requirements are more stringent than those assumed by
\cite{cypriano}. We also use a more realistic model for the wavelength
dependence of the PSF. Importantly, we examine how well the supporting
broad band imaging data need to be calibrated, as zero-point
variations will lead to coherent biases in the inferred PSF sizes. The
photometric data are also used to determine photometric redshifts, and
as a result we expect covariance between photometric redshift errors
and the inferred PSF size. The break down presented in \cite{cropper}
ignores such interdependencies, and here we examine the validity of
this assumption.

The outline of this paper is as follows. In \S\ref{theory} we present
the problem and describe the simulations we use to study the impact of
the wavelength dependence of the PSF.  In \S\ref{photoz} we explore
how well we can determine the PSF size using a conventional photometric
redshift method, whereas we investigate machine learning techniques in
\S\ref{learning}. In \S\ref{calibration} we quantify the impact of
calibration errors and limited SED templates. Appendix 
\ref{app_zband} investigates the implication of omitting \bi{z}-band
observations.

\section{Description of the problem}
\label{theory}
\subsection{Effective PSF size}

To infer cosmological parameters from the lensing data we need to
measure the correlations in the shapes of galaxies before
they were modified by instrumental and atmospheric effects. In the
following we ignore detector effects, such as charge transfer
inefficiency, which have been studied separately
\citep[e.g.][]{Massey14, Israel15}. Instead we examine how well we can
estimate the size of the effective PSF given available broad-band
observations.

We start by defining the nomenclature and notation. Throughout the
paper we implicitly assume that measurements are done on images
produced by a photon counting device, such as a charge-coupled device
(CCD). In this case the observed (photon) surface brightness or image
at a wavelength $\lambda$, $I({\bf x};\lambda)$, is related to the
intensity $S ({\bf x};\lambda)$ through
$I({\bf x};\lambda)=\lambda S ({\bf x};\lambda)T(\lambda)$, where
$T(\lambda)$ is the normalised transmission. For the results presented
here we assume that the {\it Euclid} VIS filter has a uniform transmission
between 5500-9200\AA, and that all the light is blocked at other
wavelengths. The image of an object is then given by

\be I^{\rm obs}({\bf x})=\int I^{\rm 0}({\bf x};\lambda)\otimes P({\bf
  x};\lambda){\rm d}\lambda, \ee

\noindent where $P({\bf x};\lambda)$ is the wavelength-dependent PSF,
and $I^{\rm 0}$ is the image of the object before
convolution. Following \cite{massey13}, we use unweighted quadrupole
moments $Q_{ij}$, which are defined as

\be Q_{ij}=\frac{1}{F}\int {\rm d}\lambda \int x_i x_j I({\bf
  x};\lambda) {\rm d}^2{\bf x}, \ee

\noindent where $F$ is the total observed photon flux of an image
$I({\bf x})$.

The moments can be used to estimate the shape and size of an object. A
complication is that the observed moments are measured from the noisy
PSF-convolved images, and the challenge for weak lensing algorithms is
to relate these to the unweighted quadrupole moments of the true
galaxy surface brightness distribution.  Throughout the paper we
assume that this is possible, and thus that we can use the fact that
the unweighted quadrupole moments ($Q_{ij}^{\rm 0}$) are related to the observed
quantities ($Q_{ij}^{\rm obs}$) through \citep{semboloni_colorgrad}:

\be
Q_{ij}^{\rm 0}=Q_{ij}^{\rm obs}-\frac{1}{F}\int F(\lambda)P_{ij}(\lambda){\rm d}\lambda,
\ee

\noindent where $F(\lambda)\equiv \lambda S(\lambda)T(\lambda)$
explicitly indicates the wavelength dependence of the observed photon
flux in terms of the transmission $T(\lambda)$ and $S(\lambda$), the
spectral energy distribution (SED) of the object.  $P_{ij}(\lambda)$
are the quadrupole moments of the PSF as a function of wavelength. The
second term defines the quadrupole moments of the effective PSF and
the main focus of this paper is to quantify the bias in the
measurements of galaxy shapes that arise from the limited knowledge of
the galaxy SEDs. Clearly errors in the PSF model itself contribute as
well, but we assume that these are determined sufficiently well
\citep[see e.g.][]{cropper}, although we briefly return to this in
\S\ref{calibration}.

The main complication for weak lensing measurements is that the
estimate for the effective PSF depends on the rest-frame SED of the
galaxy and its redshift, whilst neither are known a
priori. Importantly, given the large number of sources that need to be
observed to reduce the statistical uncertainties due to shape noise,
only broad-band photometry is available to estimate the SEDs and
photometric redshifts. The aim of this paper is to quantify whether
this limited information is sufficient for the accuracy we require in
the case of {\it Euclid}.

The biases in shape measurement algorithms are commonly quantified by
relating the inferred shear $\gamma$ (or ellipticity) to the true value
\citep{Heymans06}

\be
\gamma^{\rm obs}=(1+m)\gamma^{\rm true}+c,
\ee

\noindent where $m$ is the multiplicative bias and $c$ the additive
bias. \cite{massey13} examined the various terms that contribute,
including errors in the PSF model. PSF errors will lead to both
additive and multiplicative biases, although the former can be studied
from the data themselves \citep[e.g.][]{heymans2012}. Relevant here is
the bias in the estimate of the effective PSF size. We define the size
of the PSF in terms of the quadrupole moments as:

\be R_{\rm PSF}^2(\lambda)=P_{11}(\lambda)+P_{22}(\lambda), \ee

\noindent where $\rpsf(\lambda)$ is the size\footnote{Throughout the paper we refer to 
this definition as size, but note here that it  really corresponds to an area.} of the wavelength
dependent PSF, which we assume to be described by a power law

\be
\rpsf(\lambda) \propto \lambda^{0.55},
\label{r2def}
\ee

\noindent where the value of the slope is found to be a good fit to
results from simulated {\it Euclid} PSF models. In principle the wavelength
dependence can be predicted from a physical model of the optical
system, or it can be determined from careful modelling of calibration
observations of star fields. The PSF modelling greatly benefits from
the fact that stellar SEDs are well-known and well-behaved. The
observed effective PSF size $R^2_{\rm PSF}$ is then given by

\be
R^2_{\rm PSF} = \frac{1}{F}\int {\rm d}\lambda F(\lambda) \rpsf(\lambda).
\label{r2calc}
\ee

\begin{figure}
\xfigure{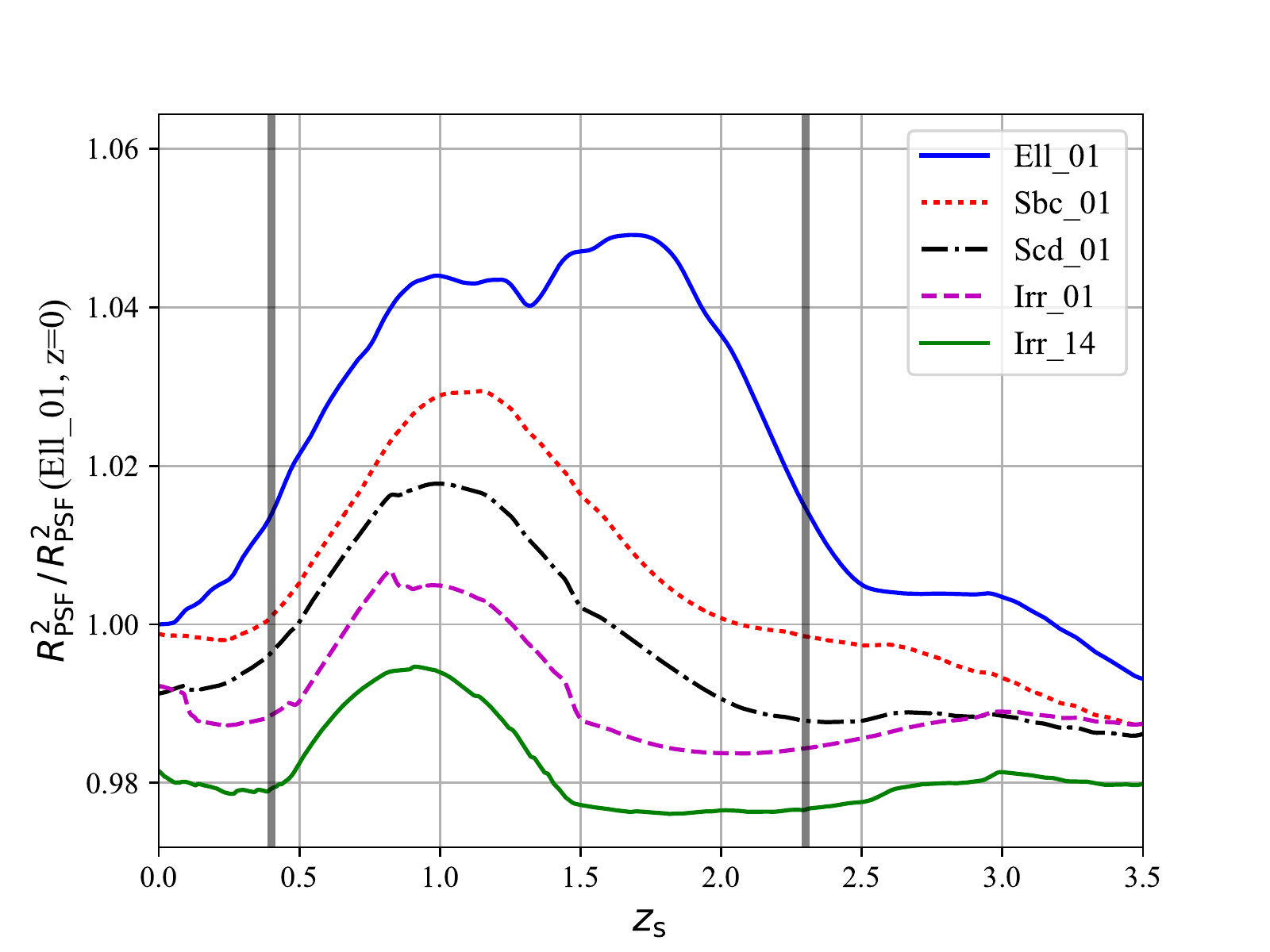}
\caption{The change in $R_{\rm PSF}^2$ as a function of redshift for
  different galaxy types, normalised to an early type (Ell\_01)
  spectrum at $z=0$. The six lines corresponds to the templates
  Irr\_01, Irr\_14, Ell\_01, Scd\_01, Sbc\_01 and I99\_05Gy from the
  CWW template library. The two vertical lines at z=0.4,1.3 indicate
  where the $\lambda = 4000$\AA\ break is entering and leaving the VIS
  filter.}
\label{r2_zdep}
\xef

\cite{cropper} presented a detailed breakdown of the various
systematic effects based on the expected performance of {\it
  Euclid}. It includes an allocation with the description `wavelength
variation of PSF contribution' in their Table~1, for which a
contribution to the relative bias in effective PSF size of
$|\delta R^2_{\rm PSF}/R^2_{\rm PSF}|=3.5\times 10^{-4}$ is listed.  To
include margin for additional uncertainties, we adopt here a slightly
more stringent requirement of

\be \left|\frac{\delta R_{\rm PSF}^2}{R^2_{\rm PSF}}\right| \equiv \left | \left
    <\frac{R^2_{\text{Pred}} - R^2_{\rm PSF}}{R^2_\text{PSF}}
  \right> \right |< \euclidreq,
\label{r2requirement}
\ee

\noindent where we explicitly average over an ensemble of galaxies
($\left< . \right>$).  The predicted value of the effective PSF size,
$R^2_{\text{Pred}}$, is the one we will attempt to estimate using
supporting broad-band observations in multiple passbands, whereas the
correct value is given by $R^2_{\text{PSF}}$. Note that this
requirement is considerably tighter than what was studied in
\cite{cypriano}.

This requirement is to be contrasted with the expected variation in
effective PSF size for different galaxy types and
redshifts. Figure~\ref{r2_zdep} shows the relative change in
$R^2_{\rm PSF}$ for five different galaxy SEDs as a function of
redshift. Both the variation between galaxy types at a given redshift,
and the variation with redshift for a given SED template are about two
orders of magnitude larger than the requirement given by
Eqn.~\ref{r2requirement}. This figure highlights that incorrect
estimates of the spectral type or photometric redshift can result in
considerable biases in the adopted effective PSF.  To explore this
problem in detail we create simulated multi-wavelength catalogs, which
we discuss next.

\subsection{Simulated data}
\label{subsec:pzmocks}

To quantify how well the effective PSF size can be determined from
broad-band imaging data, we create simulated catalogs. In this paper
we explore several approaches, which use observations of galaxies and
stars. In our forecasts we consider the combination of {\it Euclid}
observations in the VIS and NIR filters, with ground-based DES
data. This is the baseline discussed in \cite{euclidrb} and also used
in \cite{cypriano}.  We use the extended CWW library \citeind{cww} from
{\tt LePHARE} \citeind{Ilbert06} which contains 66 SEDs. We  split these into 
elliptical (Ell), spiral (Sp) and irregular (Irr) galaxies as described  in \citedir{polpz}; 
they also describes how the SEDs are assigned. These galaxy realisations with 
absolute magnitudes, redshifts and SEDs are then converted into apparent photon 
fluxes ($f_i$):

\begin{align}
f_i &= \int d\lambda \, \lambda T_i(\lambda) S(\lambda (1+z)) 
\end{align}

\xbf
\zfigure{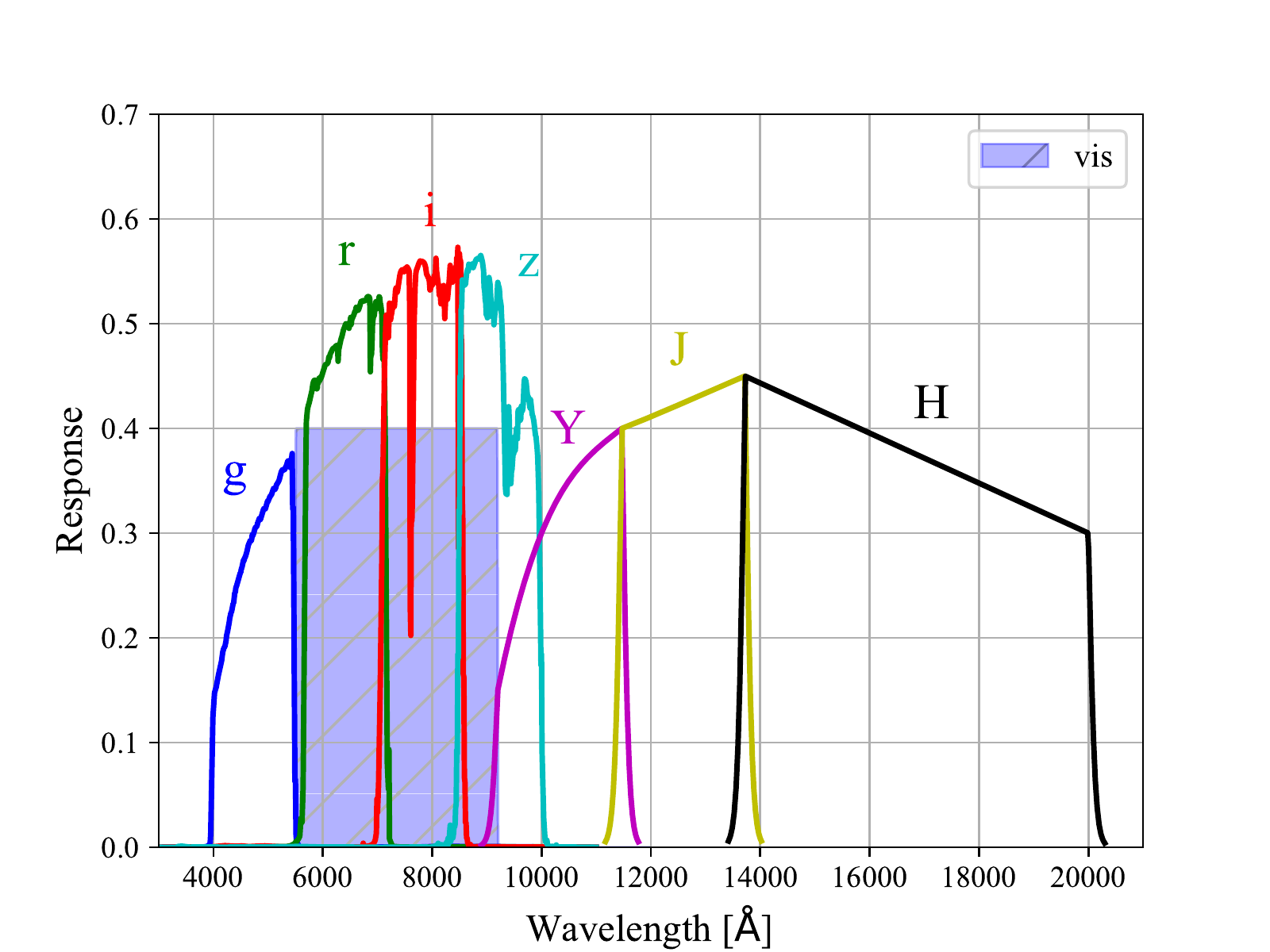}
\caption{The {\it Euclid} (VIS, and near infrared \bi{Y},\bi{J},\bi{H}
  bands) and DES (ground based \bi{g},\bi{r},\bi{i},\bi{z}) effective
  filter response curves used to create simulate data. The effective
  filter response curves combine the atmospheric (for DES), telescope,
  filter and CCD transmission.}
\label{filter:res}
\xef

\noindent
where $S(\lambda)$ is the rest-frame galaxy SED, $T_i(\lambda)$ is the
response function in filter $i$ and the integration is over the
observed frame wavelength. Figure~\ref{filter:res} shows the adopted filter
response functions ($T_i$) for the {\it Euclid} VIS and NIR filters, as well
as the optical DES filters\footnote{The DES filter curves are
  obtained from http://www.ctio.noao.edu/, while the {\it Euclid} filters
  are approximated from the values presented in \citet{euclidrb}.}.

The simulated catalogs include realistic statistical uncertainties for
the magnitudes. We assume that the measurements are limited by the
noise from the sky background, which is independent between
filters. Table \ref{mag_lim} lists the adopted limiting magnitudes for
DES and {\it Euclid} for point sources with a signal-to-noise ratio
S/N=10. For galaxies, which are extended, we take a limit 0.7
magnitude brighter. We generate mock catalogs of galaxies that include
galaxies that are fainter than $\text{VIS} < 24.5$, but restrict
the analysis to this limiting magnitude when estimating the relative
bias in $R_{\rm PSF}^2$.

We explore various scenarios to estimate the effective PSF size, such as
different combinations of broad-band imaging data.  Of particular
interest is the question whether it is possible to use the observed
sizes and colours of stars to estimate the effective PSF sizes of
galaxies: if a star and a galaxy would have the same SED, they would
also have the same effective PSF. Figure \ref{gal_star_triangle} shows
$R^2_{\rm PSF}$ as a function of colour for simulated galaxies and
stars.  The galaxies (yellow points) are a random subset of the
simulated (noiseless) catalog, while the stars (black points) are
generated by uniformly sampling all SEDs in the Pickles library
\citeind{pickles}. For the filters that overlap in wavelength with the
VIS band (panels with blue background) the relation between
$R^2_{\rm PSF}$ and colour is indeed quite similar for galaxies and
stars. The performance of this approach, extending the single colour
estimate explored by \cite{cypriano} to include the more extensive
colour information, is explored in \S\ref{learning} using machine
learning algorithms.

\begin{table}
\begin{center}
\setlength{\tabcolsep}{4.7pt}
\begin{tabularx}{\columnwidth}{lrrrrrrrr}
\toprule
{} &  VIS &    $g$ &   $ r$ &    $i$ &    $z$ &   $ Y$ &    $J$ &    $H$ \\
\midrule
Galaxies & 24.5 & 24.4 & 24.1 & 24.1 & 23.7 & 23.2 & 23.2 & 23.2 \\
Stars    & 25.2 & 25.1 & 24.8 & 24.8 & 24.4 & 23.9 & 23.9 & 23.9 \\
\bottomrule
\end{tabularx}

\end{center}
\caption{The adopted limiting magnitudes for detections with a 
  signal-to-noise ratio $S/N = 10$. The limiting magnitudes for extended 
  objects are assumed to be 0.7 magnitudes  shallower than for point sources. 
  The ground based observations (\bi{g},\bi{r},\bi{i},\bi{z}) correspond to 
  DES data, whereas the VIS,  and \bi{Y},\bi{J},\bi{H} correspond to 
  the {\it Euclid} optical and NIR limits taken from \citet{euclidrb}.}
\label{mag_lim}
\end{table}

\begin{figure*}
\yfigure{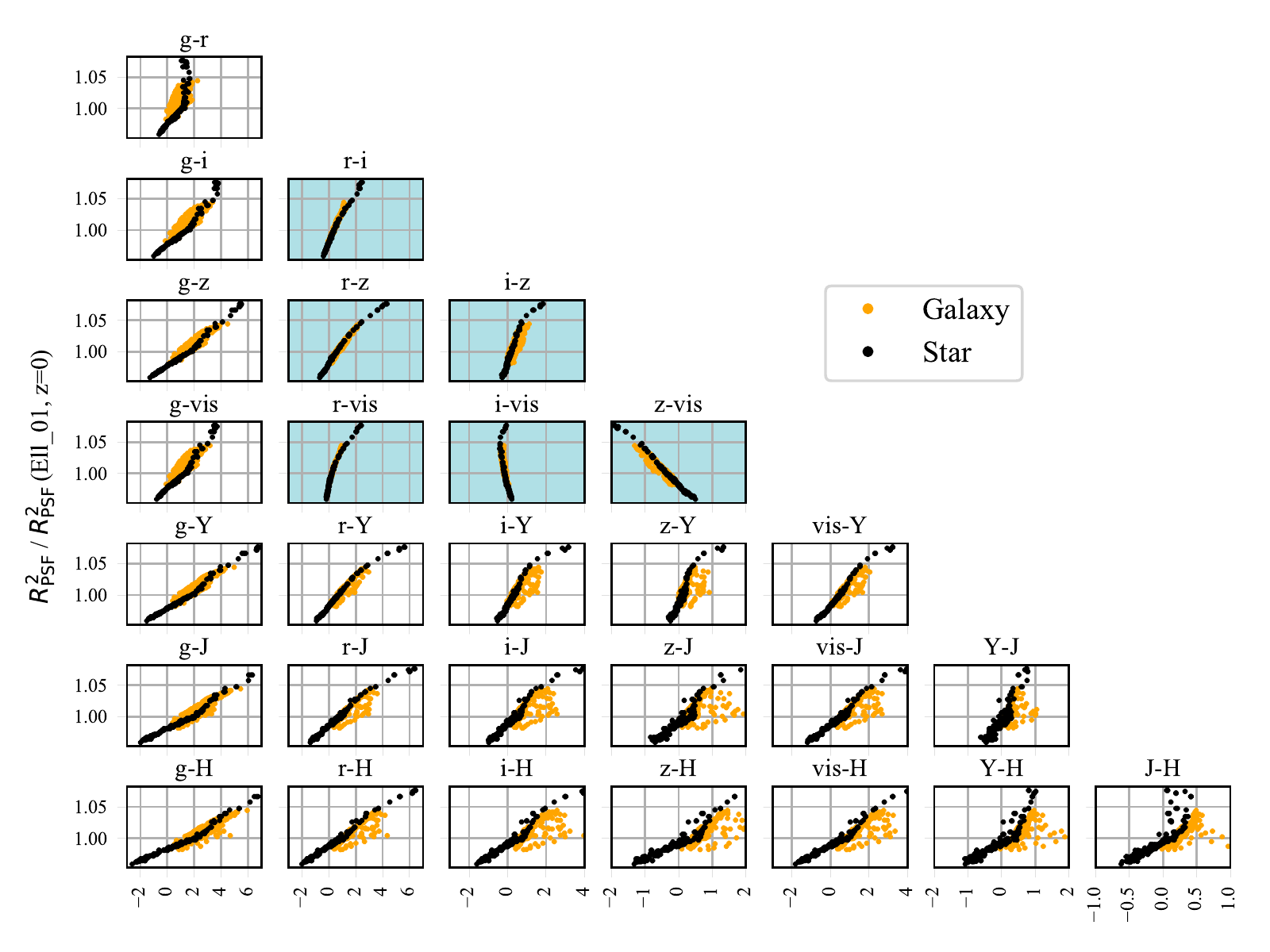}
\caption{Relation between colour and the effective PSF size ($R^2_{\rm PSF}$)
  for different filter combinations (indicated in the title of each subplot). The
  black points show the results for stars
  with SEDs from the Pickles library \citeind{pickles}. The yellow
  points correspond to a random subset of galaxies covering a range of SEDs and
  redshifts. Subplots with a blue background indicate filters
  overlapping the VIS band.  }
\label{gal_star_triangle}
\end{figure*}

\section{Performance of template fitting methods}
\label{photoz}
\subsection{Photo-$z$ estimation}
\label{subsec:pz_estim}
Cosmic shear studies rely on photometric redshifts (photo-$z$s) derived from
deep broad-band imaging to relate the lensing signal to the underlying
cosmological model. In this section we explore whether the algorithms
used to determine photo-$z$s can also be used to estimate the size of
the effective PSF.

These algorithms can broadly be divided into two classes. Machine
learning methods train on a set of galaxies where the redshift is
known from spectroscopy to predict the redshift for a larger ensemble
of galaxies only observed using broad-band photometry.  Examples of
learning methods include the neural network algorithms {\tt ANNz}
\citep{annz} and {\tt Skynet} \citep{skynet}. Template based photo-$z$
methods \citep[e.g. {\tt BPZ};][]{benitez2000,coe2006} use libraries of the
restframe galaxy SEDs. For each redshift and galaxy type, one can
model the observed galaxy colours, and the best fit model is found by
minimizing

\be \chi^2(z,\tau) = \sum_i \left[ \frac{{\left(\tilde{f}_i -
        f_i(z,\tau)\right)}^2}{\sigma_{\tilde{f}_i}^2} \right] + \
\chi^2_{\text{Priors}(z,\tau)},
\label{thr_photoz_chi2}
\ee

\noindent
after marginalizing (summing) over the galaxy types ($\tau$).  Here
$\tilde{f_i}$ and $f_i(z,\tau)$ are the observed and predicted fluxes
in the $i$th filter, with $z$ and $\tau$ being the galaxy redshift and
template, respectively.  The uncertainty in the flux is assumed to be
Gaussian with a standard deviation $\sigma_{\tilde{f}_i}$.  An
optional prior term $\chi^2_{\text{Priors}}(z,\tau)$ adjusts the
probabilities based on galaxy redshift and type, to reflect additional
constraints, for instance the fact that bright galaxies are more
likely to be at low redshift.  This term reduces the number of
degenerate solutions and catastrophic photo-$z$ outliers. In this paper
we use the default {\tt BPZ} priors specified in \citedir{benitez2000}.
The template library is based on the same set of templates used to 
create the simulations.

Unlike learning methods, template based photo-$z$ methods also provide
an estimate of the galaxy SED. From the best fit galaxy restframe SED
($S_{\rm best}(\lambda)$) and photometric redshift ($z_{\rm best}$),
one can estimate the effective PSF size

\be R^2_{\tau,{\rm best}} = \frac{\int d\lambda
  \lambda T(\lambda) S_{\rm best} (\lambda (1+z_{\rm best}))
  \rpsf(\lambda)} {\int d\lambda \lambda T(\lambda) S_{\rm
    best}\left(\lambda (1+z_{\rm best}) \right)},
\label{r2pz}
\ee

\noindent
where the integration is over the observed-frame wavelength. However,
because of measurement uncertainties in the photometry, the template
based codes not only provide the best fit redshift for each galaxy,
but also a redshift probability distribution

\be
p(z) \propto \exp{\left( - \frac{1}{2} \sum_\tau \chi^2(z, \tau) \right)},
\ee

\noindent where $\chi^2$ is given by Eq.\ref{thr_photoz_chi2}. Instead
of using the best fit redshift (Eq. \ref{r2pz}), one can instead
estimate a PSF size, weighted by the redshift probability distribution:

\be R^2_{\tau,p(z)} = 
\frac{\int dz p(z) \int d\lambda \lambda
  T(\lambda) S_{\rm best}\left(\lambda (1+z)\right) \rpsf(\lambda)}
{\int dz p(z)\int d\lambda \lambda T(\lambda) S_{\rm best}\left(\lambda (1+z)
  \right)}.
\label{r2pdf}
\ee

\noindent This can be extended further by also including the
probabilities of the galaxy types.  We compare the performance of
these choices for the effective PSF size estimates in
\S\ref{subsec:pzconv}.

The template library uses the following six templates: Ell\_01,
Sbc\_01, Scd\_01, Irr\_01, Irr\_14, I99\_05Gy, which represent a
subset of the SEDs in the galaxy mocks. This limited set of SEDs
reflects the conventional use of photo-$z$ algorithms and the fact that
the real SEDs are not perfectly known. We explore this in more detail
in \S\ref{subsec:sedfit}, but note that the photo-$z$ code does
include two linear interpolation steps between consecutive templates
to mimic a smooth transition between templates. When determining
photometric redshifts from the mock galaxy catalogs we exclude the VIS
band, since it only yields a minor improvement in the photo-$z$
precision, while it would increase the covariance between the
photometric redshift and the shape measurements.

\subsection{Simple scenario}
\label{subsec:pzgauss}

The photo-$z$ algorithm provides an estimate of the restframe SED, while the
calculation of the effective PSF size is done using the galaxy SED in
the observed frame.  The conversion between the two frames requires
the redshift, which causes the redshift biases and uncertainties to
directly affect the estimate of $R^2_{\rm PSF}$ from a template based
photo-$z$ code. The photo-$z$ probability density distribution that is
provided by a template fitting algorithm can be a complex function of
redshift, as degenerate solutions may be found with different best-fit
SEDs. Before examining this complex, but more realistic situation, we
consider a number of simpler cases that allow us to disentangle the
different effects.

\xbf
\xfigure{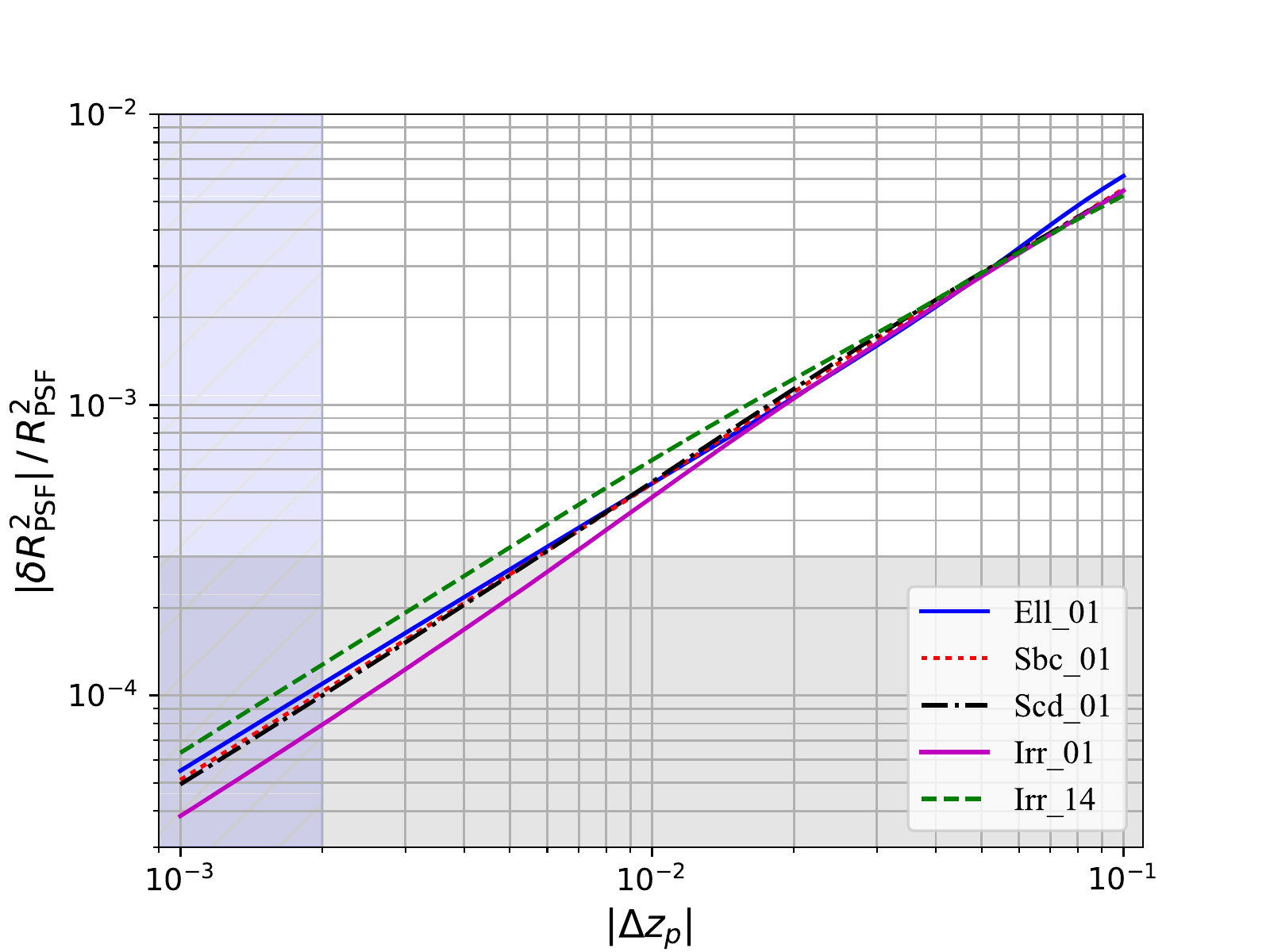}
\xfigure{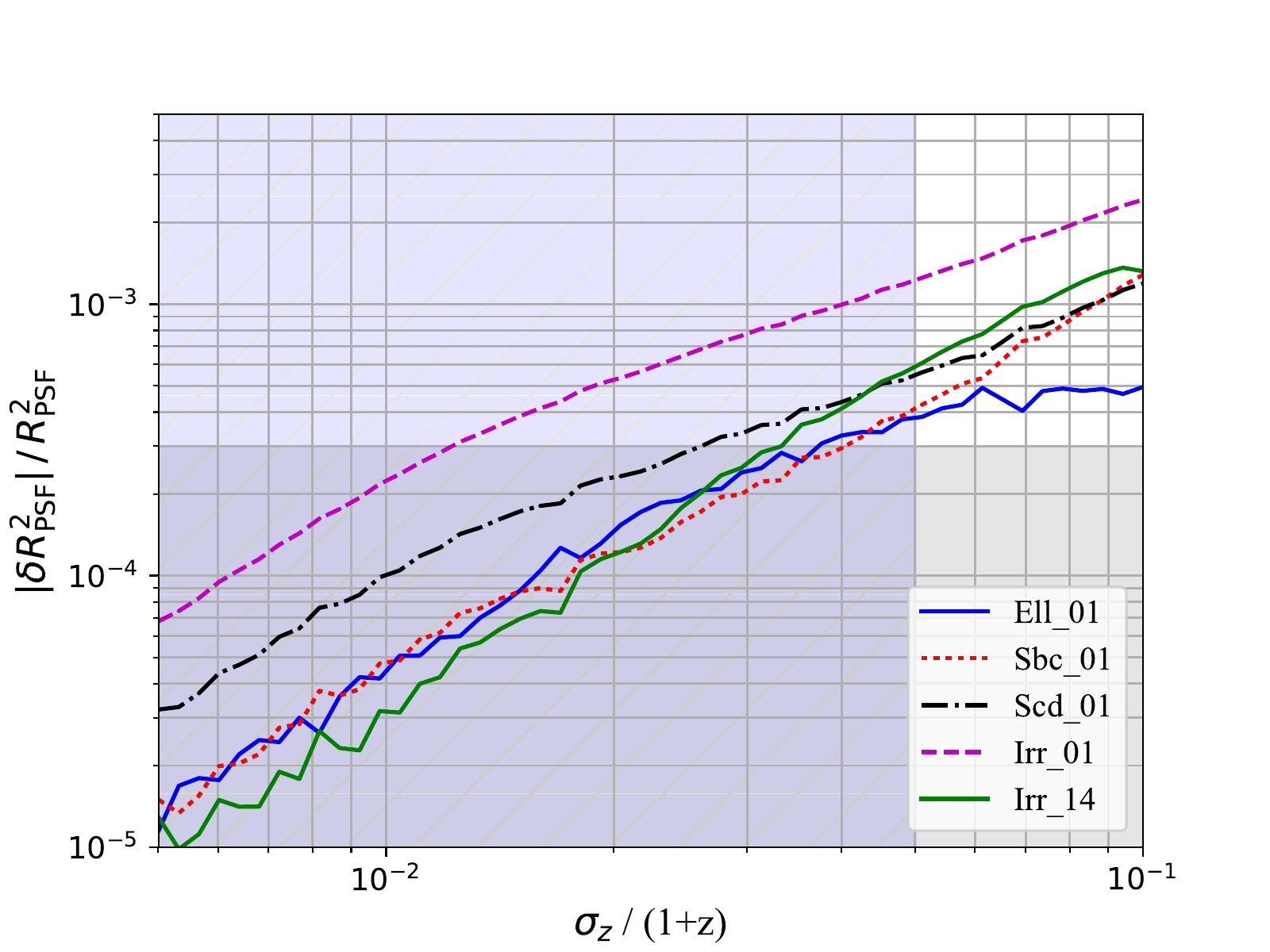}
\caption{Effect of Gaussian photo-$z$ uncertainties on the estimate of
  $R^2_{\rm PSF}$ for a known rest-frame SED. The lines show the
  relative bias $\delta R^2_{\rm PSF}/R^2_{\rm PSF}$ for different
  galaxy templates at $z = 0.5$. In the top panel the photo-$z$ bias
  varies (no photo-$z$ scatter), while the bottom panel varies the
  photo-$z$ scatter (no photo-$z$ bias).  The solid (gray) shaded region
  shows the required accuracy. In the top panel the hatched (blue)
  region indicates a photo-$z$ bias $<0.002(1+z)$ within a redshift bin,
  while it marks a photo-$z$ scatter $<0.05(1+z)$ in the bottom panel.}
\label{r2_gauss}
\xef

The top panel in Fig.~\ref{r2_gauss} shows the absolute value of the 
relative bias in effective PSF size if we assume that the SED is known a 
priori, but where the best-fit photo-$z$ is biased by $|\Delta z_p|$.  While the size
of the bias varies with redshift, we limit the discussion here to
$z=0.5$ for simplicity, and use the full redshift range for the
realistic simulations (see \S\ref{subsec:pzconv}). The amplitude of
the bias increases with increasing redshift bias, with the different
templates yielding rather similar results.  We find that if
$|\Delta z_p|<0.005$ the resulting bias in PSF size is within the
adopted allocation for {\it Euclid} (indicated by the grey shaded
region). This may appear challenging, but a correct interpretation of
the cosmic shear signal requires that the bias in the mean redshift
for a given tomographic bin is known to better than
$|\Delta z|<0.002(1+z)$ \citep{euclidrb} and thus for an ensemble of
galaxies the resulting bias in the PSF size may be sufficiently
small. However, the redshift sampling of photometric redshift codes
is typically $\Delta z\sim 0.01$, which may introduce biases. We examine
this in detail in Appendix~\ref{app_dzresol} and find this is not a concern.

The situation is more problematic when we consider the uncertainty in
the photometric redshift estimate, which we assume to be a Gaussian
with a dispersion $\sigma_z$ around the correct redshift (i.e. no
bias).  The bottom panel in Fig.~\ref{r2_gauss} shows the relative
bias in the effective PSF size as a function of $\sigma_z/(1+z)$ for
galaxies at $z=0.5$, demonstrating that a small photo-$z$ scatter can
cause a substantial bias in the estimate of the effective PSF
size. The bias increases with increasing uncertainty, with the largest
bias occurring for the Irr\_01 template.  The requirements for the
cosmic shear tomography for {\it Euclid} are that
$\sigma_z/(1+z)<0.05$ \citep{euclidrb}, and hence the resulting biases
in PSF size are somewhat larger than can be tolerated.  However, in
reality one averages over a sample of galaxies within a tomographic
bin, and thus these numbers should not be considered appropriate
requirements. Nonetheless they indicate that the statistical
uncertainties in the photometric redshifts are important.

\subsection{Conventional template fitting method}
\label{subsec:pzconv}

After considering the simplistic case of galaxies with a known SED and
Gaussian photo-$z$ errors, we now examine the performance
of a template fitting method using the more realistic galaxy
simulations described in \S\ref{subsec:pzmocks}. As a consequence, the
results include redshift outliers and misestimates of the galaxy
rest-frame SED.

\begin{table}
\begin{center}
\begin{tabular}{ l c c c c }
\hline

\input{prod_psf43.txt}

\hline
\end{tabular}
\end{center}
\caption{The mean relative bias in effective PSF size times $10^{-4}$ when using
  a photo-$z$ template fitting method. Columns show the results
  by galaxy type: Elliptical (Ell), Spiral (Sp) or Irregular (Irr) galaxies.
  The rows list the relative bias when using the best fit photo-$z$ (Peak) and
  when weighting using the redshift pdf (PDF).}
\label{tbl_pzbias}
\end{table}

\xbf
\xfigure{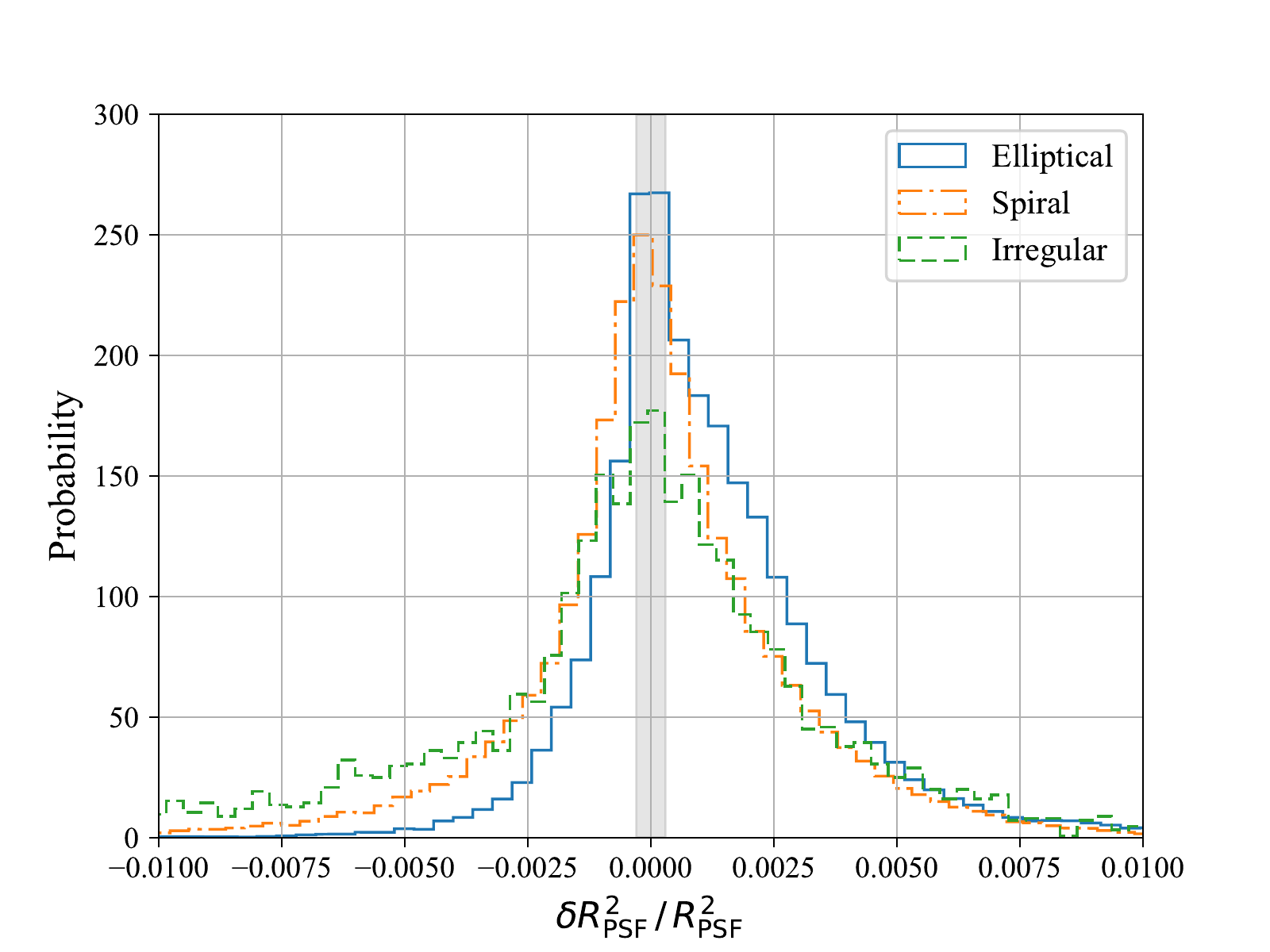}
\caption{The distribution of relative biases
  $\delta R^2_{\rm PSF}/R^2_{\rm PSF}$ for different galaxy types
  using a photometric redshift template fitting code. The galaxies are
  split into Elliptical, Spiral and Irregular types based on
  the definition in the input mock catalogue. The vertical band marks
  the {\it Euclid} requirement for the mean relative bias.}
\label{r2pz_hist}
\xef

As mentioned earlier, the template fitting algorithm provides an
estimate for the best fit redshift and rest-frame SED, but also a
probability density distribution for the redshift (which may be
combined with a distribution of templates $\tau$). The different
outputs can be used to estimate the average bias in the effective PSF
size. Table~\ref{tbl_pzbias} lists the resulting average values for
$\delta R^2_{\rm PSF}/R^2_{\rm PSF}$ when splitting by the true galaxy
types. If we consider the best fit redshift estimate, the biases are
small for both the spiral and irregular galaxies, whereas the biases
are large for early type galaxies, irrespective of the weighting
scheme.

It is also instructive to examine the distribution of biases for the
different galaxy types. Figure~\ref{r2pz_hist} shows that the
distribution of effective PSF sizes is much broader than the
requirement. For all three galaxy types the distribution peaks close to zero, 
and the bias for all three types is caused by the skewness towards larger $R_{\rm PSF}^2$
values, because the redshift errors and SED misestimates do not
fully cancel.

\xbf
\xfigure{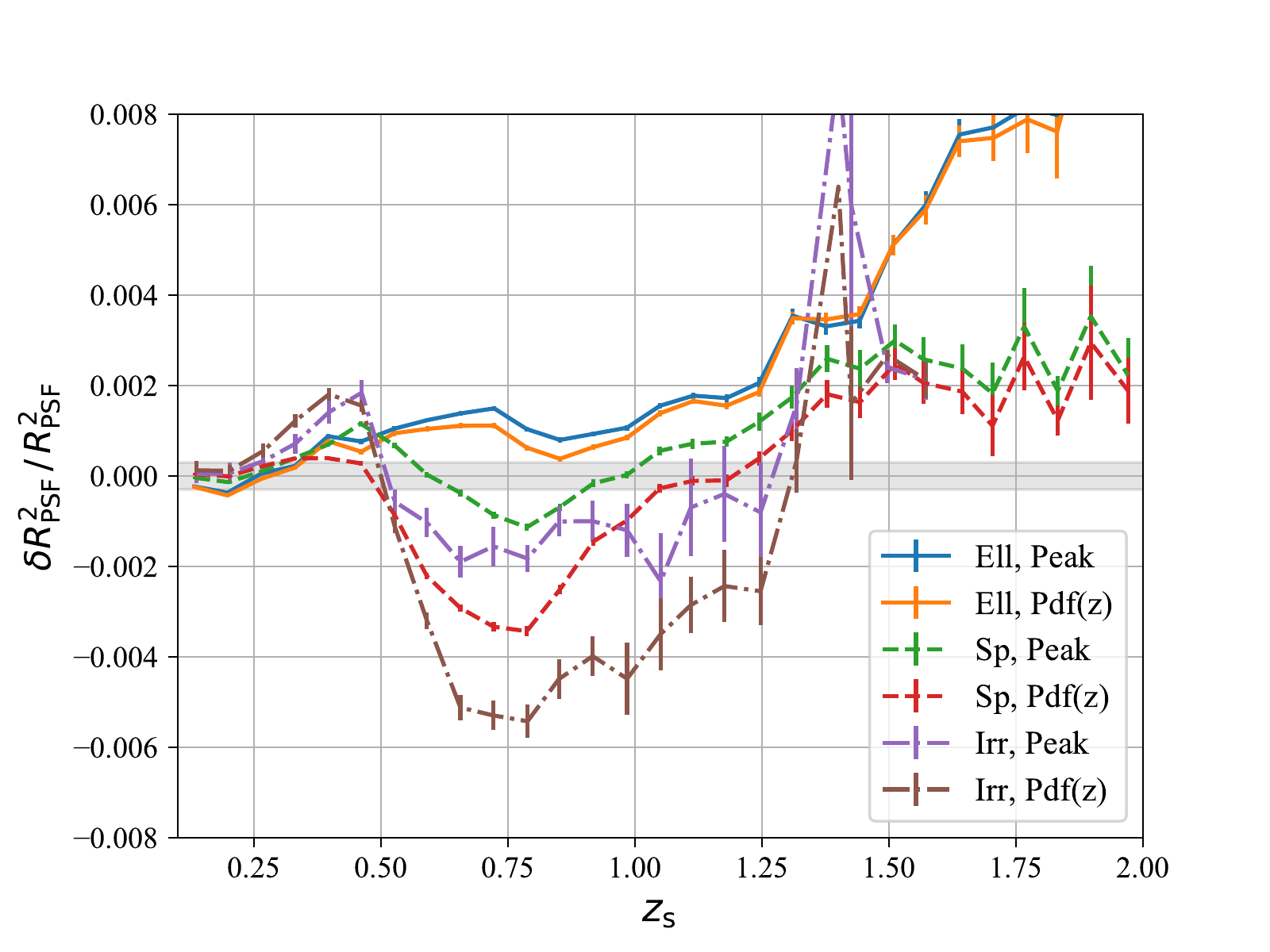}
\caption{Redshift dependence of the relative bias in the effective PSF
  size when a photo-$z$ template fitting method is used. The plot shows
  results split by input galaxy type and for two approaches to
  determine $R^2_{\rm PSF}$, i.e. using the best fit photometric
  redshift, or using the full $p(z)$. In both cases we do not
  marginalize over the uncertainty in the SED. The horizonal band
  marks the {\it Euclid} requirement.}
\label{r2pz_redshift}
\xef

The results in Table~\ref{tbl_pzbias} are averages over the full
redshift range, but the broad distributions in Fig.~\ref{r2pz_hist}
for the three galaxy types suggest that other parameters play a
role. Figure~\ref{r2pz_redshift} shows the relative bias in
$R^2_{\rm PSF}$ for different estimators as a function of redshift.
Note that the number of irregular galaxies in our simulations is
negligible at $z>1.5$, because they are fainter than our magnitude
cut, and the measurements therefore only extend to this redshift. The variation 
as a function of redshift is the main cause of the broad
distributions in Fig. ~\ref{r2pz_hist}, and is much larger than can
be tolerated.  This demonstrates that a simple average over the full
galaxy sample is not adequate.  In general using the best-fit redshift
and the redshift weighting yields very similar results, although the
averages differ somewhat. We therefore focus on the results using
the best-fit redshift below.

\subsection{Restricted template fitting}
\label{subsec:sedfit}

It is interesting to investigate which parameters of the experimental setup
affect the bias in effective PSF size the most, as they may provide clues
how to improve the performance. We therefore examine a range
of scenarios where we modify the set of filters used, examine the SED coverage
of the algorithm, as well as the role of the photo-$z$ priors.

\xbf
\xfigure{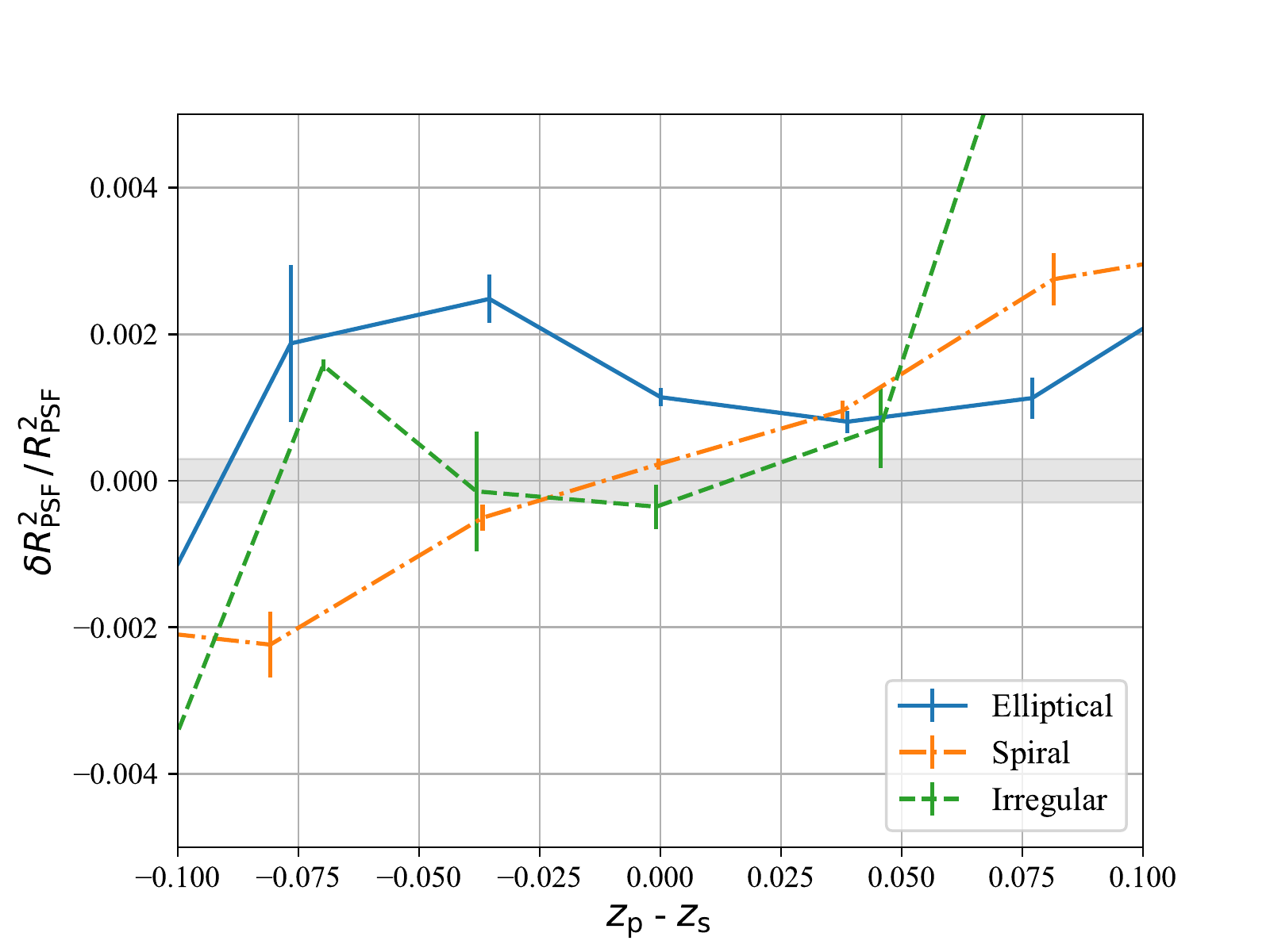}
\caption{Effect of a redshift error on the estimate for
  $R_{\rm PSF}^2$. The plot shows the relative bias
  $\delta R^2_{\rm PSF}/R^2_{\rm PSF}$ when using a photo-$z$ template
  fitting method for three input galaxy types as a function of the
  difference between estimated and true redshift. The horizonal band
  marks the {\it Euclid} requirement.}
\label{r2pz_deltaz}
\xef

The observed frame SED in the VIS filter is the most important
quantity when estimating the effective PSF size because the shapes are
measured using these images. In \S\ref{subsec:pzgauss} we already saw
that the errors in the photometric redshifts can introduce bias. As we
kept the SED fixed in this case, we effectively modified the observed
SED. One might naively assume that the algorithm will adjust the best
fit SED accordingly, thus reducing the bias. To quantify this, we show
the relative bias in effective PSF size as a function of the
difference between the best-fit photo-$z$ $z_p$ and the true redshift
$z_s$ in Fig.~\ref{r2pz_deltaz}. In this case the algorithm is free
to adjust the SED.  For the late type SEDs the results look
qualitively similar to what was found in \S\ref{subsec:pzgauss}, but
for the early type galaxies the effective PSF size is overestimated
consistently. Hence it is incorrect to assume that errors in the
photo-$z$ estimate are compensated by selecting a different SED. We
find that only using \riz\ data does not reduce the bias.

In template based photo-$z$ methods the flux measurements in the various filters
are compared to the model, and additional priors are used to restrict the range
of solutions. The former can be split further into the contributions that arise
from the \riz\ filters that overlap with the VIS band, and the out-of-band filters
($\bi{g},\bi{Y},\bi{J},\bi{H}$ in our case). Minimising $\chi^2$ using the contributions from
the out-of-band filters and the photo-$z$ priors only provides indirect information
on the SED in the VIS band and may thus lead to biases in the estimate for
the PSF size. It is therefore of interest to examine whether the performance
improves by restricting the filters used. Reducing the statistical uncertainties
in the flux measurements may provide another way to improve the performance.

\begin{figure*}
\yfigure{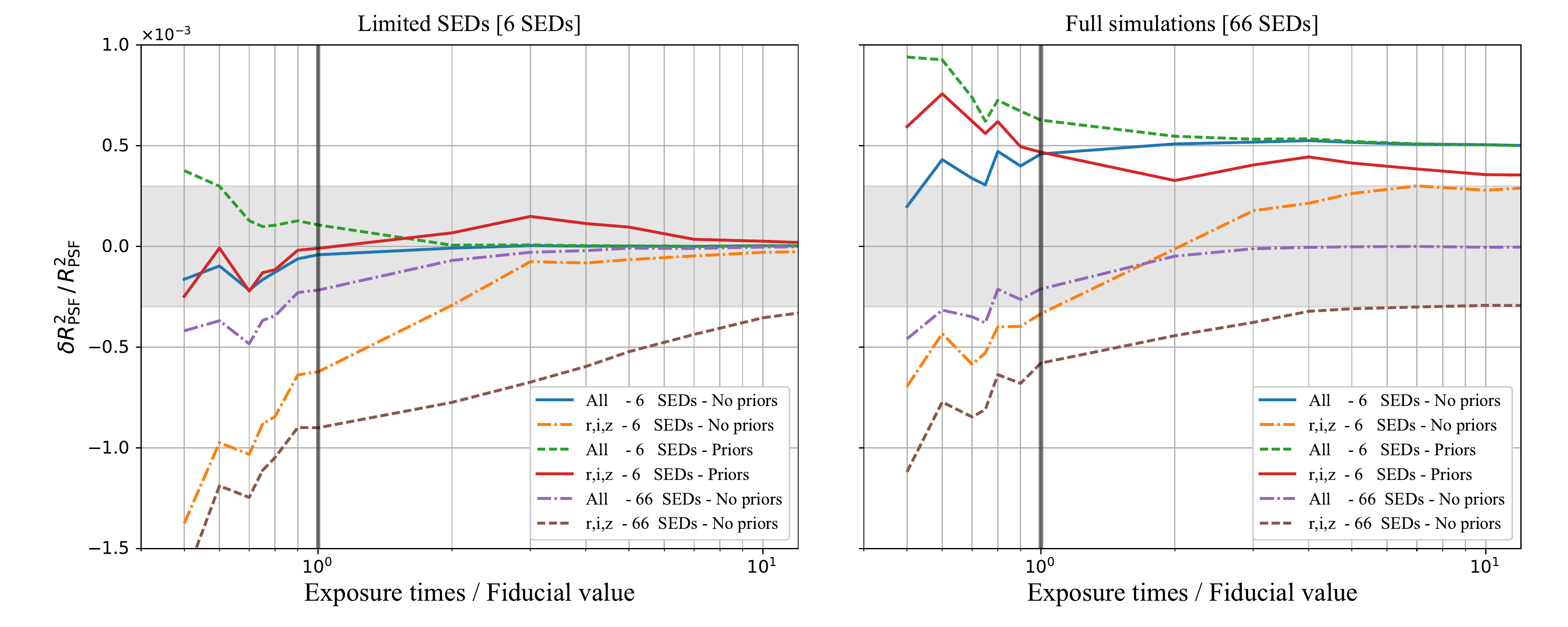}
\caption{$\delta R^2_{\rm PSF}/R^2_{\rm PSF}$ for the SED template
    fitting method. On the x-axis the exposure time is scaled relative
    to the fiducial setup. In the left panel the simulated galaxy
    catalog only include the 6 templates used in the photo-$z$ code,
    while for the right panel it includes the full 66 templates. The
    lines show which combination of filters was used (All or
    $r,i,z$), whether priors were used (No priors or Priors),
    and how many SEDs were used. The vertical line indicates
    the fiducial setup, while the shaded (gray) band marks the
    {\it Euclid} requirement.}
\label{r2sed}
\end{figure*}

Figure~\ref{r2sed} shows the relative bias for the full sample as a
function of exposure time (relative to the nominal case), where we
assumed that the increase in exposure time is the same for all
filters, including the {\it Euclid} VIS and NIR filters. We do so for
different setups. In the left panel we create simulated catalogs using
only six distinct SEDs, whereas in the right panel the full range of
SEDs is used. We do keep the luminosity functions for the various
galaxy types unchanged, but rather assign slightly different SEDs to
each type.

For the simulations with six SEDs (left panel) there are several
combinations of filters that meet the requirement on the average
relative bias (as indicated by the grey region).  If the filter set is
restricted to \riz, the priors are needed to reduce the average bias
for the fiducial exposure time, but otherwise applying the photo-$z$
code with only six templates performs well.  As expected, in the noiseless limit
the bias vanishes for all setups with six templates.  While
the galaxy priors and the \bi{g},\bi{Y},\bi{J},\bi{H} only contribute
indirectly to constrain the SED within the VIS band, including these
does reduce the bias in the PSF size. 

Of particular interest are the two cases where the photo-$z$ algorithm
uses all 66 templates in the analysis: the biases are larger, although
they do vanish in the noiseless case (with the \riz\ scenario
converging outside the plot). The larger bias can be understood,
because the noise causes the algorithm to select SEDs not in the
simulations. While adding more templates in the fitting code may be
tempting, this result cautions us that it can also lead to biases.

The right panel in Fig.~\ref{r2sed} shows the results when the full
range of SEDs is used to create the simulated catalogs. In this case
we find that using only six templates in the photo-$z$ algorithm leads
to biases, even in the noiseless case. The \riz\ setup without photo-$z$
priors accidentally meets the requirements for the nominal exposure
time. Using the full range of SEDs in the analysis improves the
performance, as expected, with the best results for the case where all
filters are used. These results highlight the need for the photo-$z$
templates to span the full set of SEDs in the observations, which may
be challenging in practice.

\section{Machine learning techniques}
\label{learning}
The results presented in the previous section suggest that modifications to the 
template fitting codes are needed if these are to be used to determine the effective PSF.
Motivated by the fact that Fig.~\ref{gal_star_triangle} shows that the PSF sizes for galaxies
correlate strongly with the observed colour, we explore the use of machine learning methods 
as a possible alternative to map between the observed colours and the effective PSF. 

Although machine learning techniques are fast, flexible and easy to implement, the spatial
variation of the PSF introduces additional complications which may be
less straightforward to implement; in contrast the effective PSF is readily
computed given a model for the PSF and the SED from a template fitting code.
Regardless of which approach may be best suited for the analysis of {\it Euclid} data, 
quantifying the performance of the machine-learning methods allows us to assess 
whether the supporting ground-based observations for {\it Euclid} are adequate to 
infer the effective PSF.

\subsection{Training on simulated galaxy photometry}
\label{subsec:basic_train}

\xbf
\xfigure{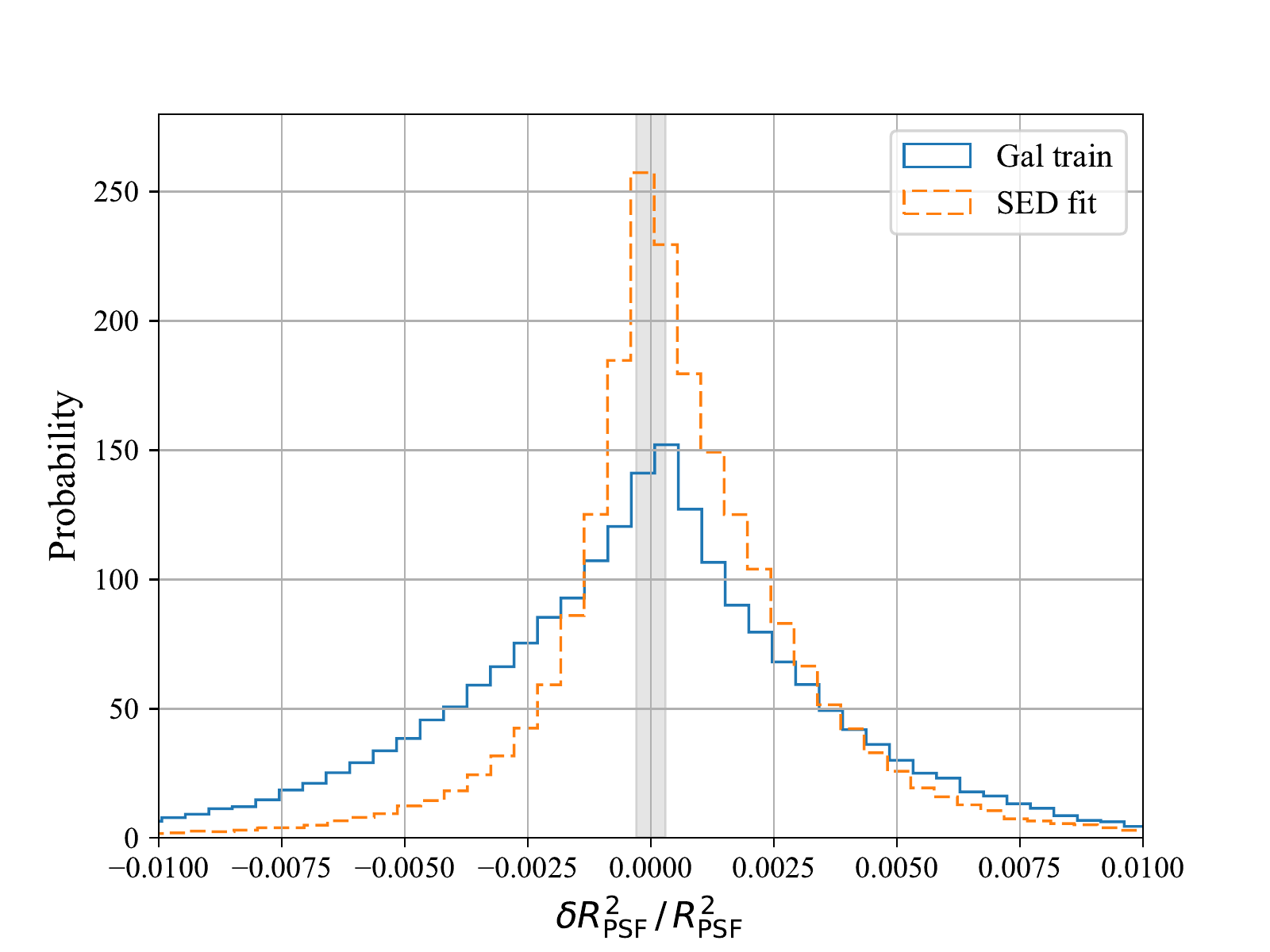}
\caption{Histogram of $\delta R^2_{\rm PSF}/R^2_{\rm PSF}$ values when
using a template fitting photo-$z$ code (dashed green) and when
training on \riz\ data from a simulated galaxy catalog (blue).}
\label{r2pz_r2train}
\xef

To explore the performance of machine learning we create a training sample of galaxy 
simulations using the same procedure as the test catalog (as described in
\S\ref{subsec:pzmocks}). We start with a best case scenario where the
training set does not contain noise. The {\tt NuSVR} algorithm from
{\tt scikit-learn} \citeind{scikitlearn} is used to train on a
sample of 4000 galaxies with multi-wavelength measurements.
The galaxy training and test catalogs are generated with
the same algorithm, but they are separate realisations. 

The results are used to estimate the effective PSF sizes of the test
catalog (which does contain noise). A histogram of the residuals
when we train on 4000 galaxies with \riz\ photometry
is presented in Fig.~\ref{r2pz_r2train} (solid histogram). The
distribution of residuals is fairly symmetric and centred around zero
bias: we observe an average value of
$\delta R_{\rm PSF}^2/R_{\rm PSF}^2=3.5\times 10^{4}$ (also see
Table~\ref{training_table}). For comparison we also show (dashed
histogram) the residuals from the template fitting method (all
filters, priors, photo-$z$ peak). The residuals from the SED fitting
method also peak around zero, but the distribution is noticeably
skewed, which leads to a significant bias when averaged over the full
sample.

We also consider a number of scenarios where different filter combinations
are used to train. The resulting average biases are listed in Table~\ref{training_table}. 
The results in the second column are for noiseless photometry, whereas
noisy data were used in the training for the results listed in the third column.
We find that all configurations perform well, with the exception of the case
where only $i,z$ data are used to train.  This can be understood from
Fig.~\ref{gal_star_triangle}, which shows that for all colours the
colour-$R^2_{\rm PSF}$ relation is tightest for colours that include the
\bi{r}-band.  Interestingly, we find that the combination of $r,i$
performs better than the \riz\ setup. This is because the VIS filter
covers only the blue half of the $z$-band. Hence the information
contained in the $z$-band measurements is of limited value as some of
the flux falls outside the VIS filter. We verified this with a test
where the $z$-band was restricted to the blue half: the bias was
reduced, whereas the bias increased when we used the redder half.

Although perhaps somewhat counterintuitive, we advocate to exclude the
$z$-band when a restricted set of filters is used to estimate the
effective PSF when training using galaxy templates. We explore this
setup in more detail in Appendix~\ref{app_zband}. Including more
filters does reduce the bias, but it is not clear whether this would
work in practice because of variations in photometric
calibration. Especially including the NIR data would lead to
additional requirements on the relative calibration between the
ground-based optical data and the {\it Euclid} NIR data.

\begin{table}
\begin{center}

\begin{tabular}{lrrrr}
\toprule
     Filters &     Gal &  Gal-noisy &    Star &  Star-noisy \\
\midrule
       r,i,z & 1.2 &    3.0 & 2.1 &     2.8 \\
         r,i & 0.4 &    1.0 & 9.7 &     12 \\
         i,z & 11 &    13 & 40 &     38 \\
   vis,r,i,z & 1.4 &    1.8 & 1.2 &     2.0 \\
     vis,r,i & 2.0 &    0.2 & 7.4 &     8.5 \\
     vis,i,z & 1.3 &    7.7 & 27 &     26 \\
         All & 0.1 &    1.7 & 9.9 &     8.3 \\
   All - vis & 0.1 &    1.8 & 1.7 &     2.5 \\
     All - r & 0.4 &    4.7 & 1.8 &     1.8 \\
     All - z & 0.1 &    1.0 & 6.1 &     4.7 \\
 All - vis,z & 0.1 &    1.8 & 4.4 &     3.7 \\
\bottomrule
\end{tabular}

\caption{
  The absolute value times $10^{4}$ of the relative bias  $|\delta R^2_{\rm PSF}|/R_{\rm PSF}^2$ when 
  using a machine learning method
  for different instrumental setups and training approaches. The first
  two columns (Gal) list results when the algorithm is trained on
  galaxy simulations, while the last two columns (Star) correspond to the
  biases when the algorithm is trained on stars. The columns marked
  `-noisy' include noise and the others are noiseless.  The first column
  lists the filters used. Here 'All' means \bi{g},\bi{r},\bi{i},\bi{z},VIS,\bi{Y},\bi{J},\bi{H}.
  and `All - filter' means that the filter was omitted.}
\label{training_table}
\end{center}
\end{table}

\subsection{Training on observed photometry of stars}
\label{subsec:star_train}

So far, our attempts to predict the effective PSF size have focused on
galaxy templates. The predicted SED is then used with a model of the
wavelength dependence of the PSF to compute the value for
$R_{\rm PSF}^2$. On the other hand, the PSF properties, including
higher order moments of the surface brightness distribution, can be measured 
directly for stars in the data.  Moreover, Fig.~\ref{gal_star_triangle} shows that 
galaxies and stars with the same colour have very similar effective PSF sizes. 
It is therefore interesting to examine whether it is possible to train on a sample of
stars instead. 

The main benefit of such an approach is the potential of constructing a self-calibrating method: 
when training on, and applying the results to the same pointing, this implementation would
be rather insensitive to calibration errors in the photometric zero-points. The effective PSF 
can be computed using the wavelength-dependent model of the PSF and the SEDs of the stars. 
The resulting mapping between effective PSF parameters and observed star colours could then 
also be applied using the observed colours of galaxies. As before, we focus on the effective PSF size
for simplicity.

We start with noise-free measurements of the colours of stars. For the
star catalogs we use 400 stars, uniformly sampled from stellar SEDs in
the Pickles \citeind{pickles} library. A typical {\it Euclid} pointing
is expected to contain more stars, and thus these results give a
conservative indication whether self-calibration is feasible. On the
other hand, the distribution of stellar SEDs is wider than in actual
data. We explore this further in \S\ref{subsec:tomo} where we use a
realistic distribution of spectral types from the second data release
of the KiloDegree Survey \citep[KiDS;][]{Kuijken15}. 

To mimic a self-calibration procedure, we create 100 independent
pointings, and train the NuSVR algorithm on the star simulations.  
The last two columns in Table~\ref{training_table} list the results
when we train using the star catalogs. The biases are generally small,
and in some (noise-free) cases this approach outperforms the training
on galaxy templates. This the consequence of the fact that the stars
show a remarkably simple relation between the colour and
$R_{\rm PSF}^2$ (see Fig.~\ref{gal_star_triangle}). An important
difference is the increase in bias when excluding the \bi{z}-band,
which did not occur when training with galaxies. Including the
VIS band reduces the bias, albeit with limited effect. As was the
case for the galaxies, including VIS and NIR data can help, provided
the relative calibrations between the various data sets can be
ensured.

\subsection{Tomography and calibration sample}
\label{subsec:tomo}
\xbf
\xfigure{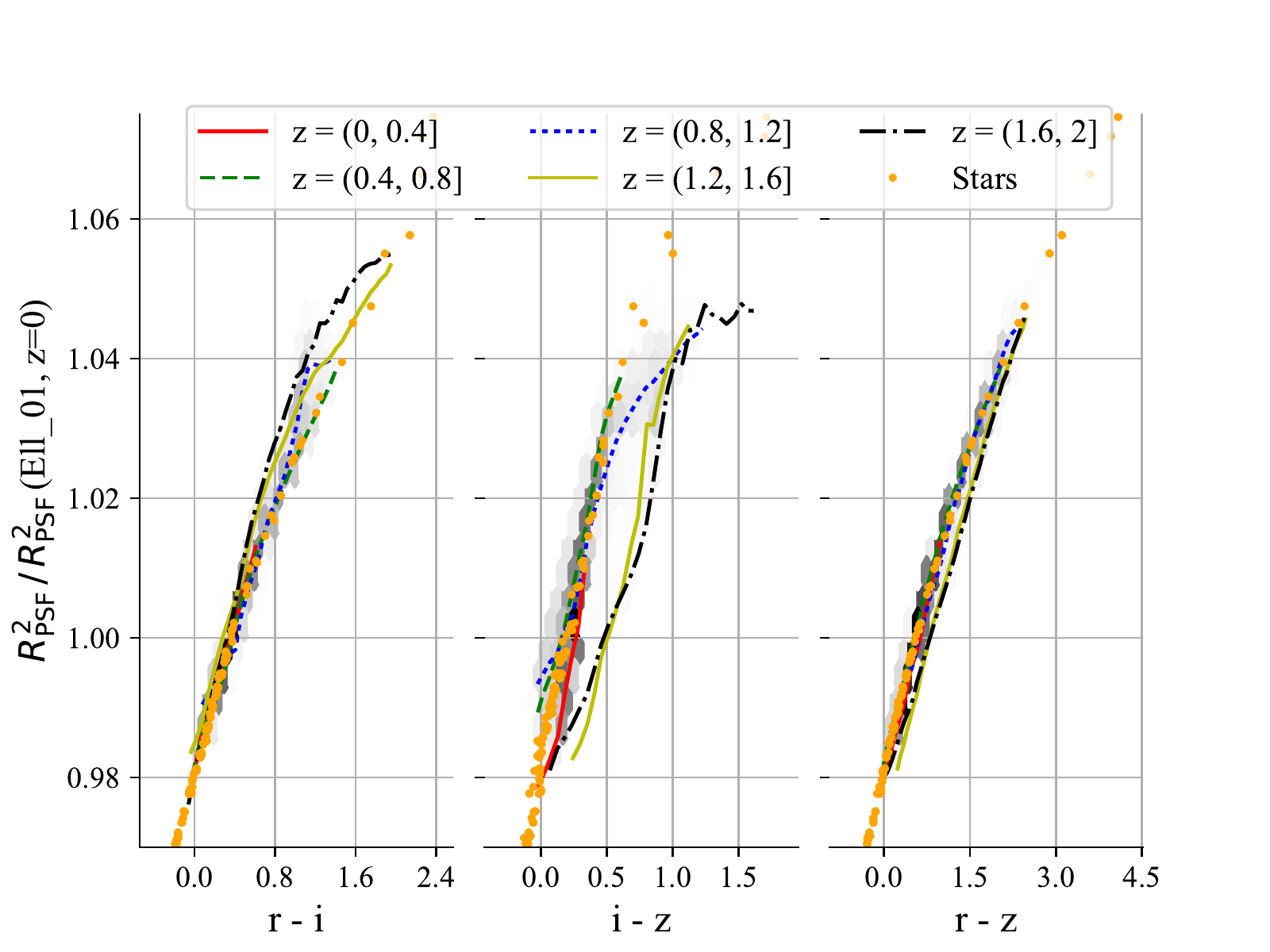}
\caption{Effective PSF size $R^2_{\rm PSF}$, relative to the value
for the Ell\_01 SED at $z=0$, versus colour for filters
that overlap with VIS. Stars are marked as dots, while galaxies
are binned in redshift ranges and shown as lines. The hex-bins shows
the density for the full galaxy population. No magnitude noise is
included.}
\label{stars_failure}
\xef

The constraints on cosmological parameters from cosmic shear surveys
are improved significantly if the source sample is split in a number
of narrow redshift bins, such that they are sensitive to the matter
distribution at different redshifts.  Such `tomographic' analyses are
now standard for cosmic shear studies \citep[e.g.][]{Heymans13, Becker16, Jee16, 
Hildebrandt17}, and therefore it is not sufficient to consider the bias for the full
sample, especially because the effective PSF size varies strongly with
redshift. Hence, even though training on stars results in an overall
small bias in the effective PSF size, we need to ensure that the bias
does not vary significantly with redshift. We already saw that this
was problematic for template fitting methods (see
Fig.~\ref{r2pz_redshift}).

Figure~\ref{stars_failure} shows the relative change in
$R^2_{\rm PSF}$ for galaxies and stars as a function of the colours
that overlap the VIS-band. Similar to what we saw in
Fig.~\ref{gal_star_triangle}, the stars (orange points) trace the
overall galaxy population well (grey hexagons), but when we split the
galaxy sample into redshift bins (indicated by the lines), we find
that this good correspondance does not hold for all redshifts:
especially for the highest redshifts the relation is very different,
which is problematic for a machine learning method. This result is
similar to the conclusion reached by \cite{cypriano} who considered
only a single colour.

\xbf
\xfigure{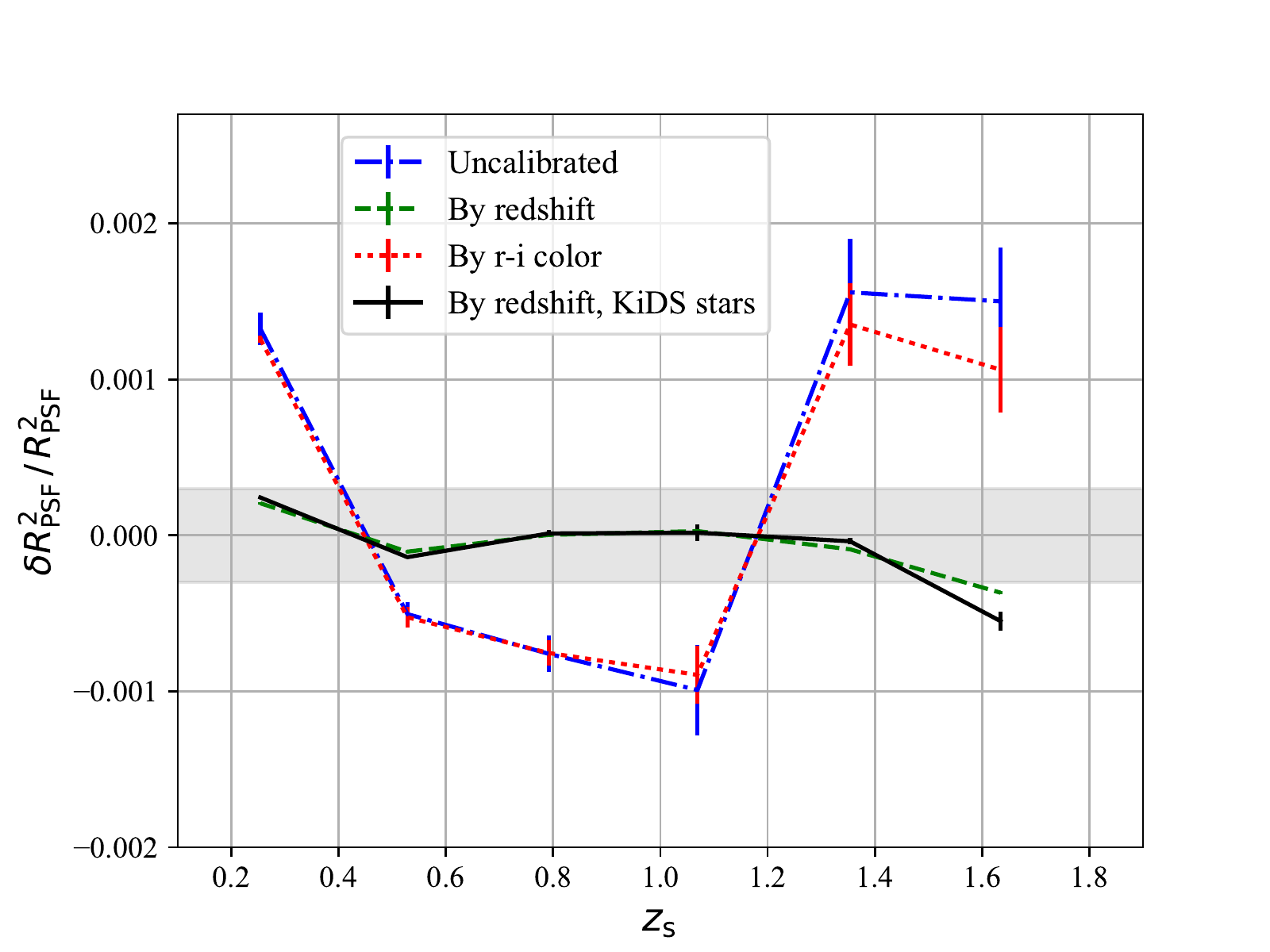}
\caption{The relative bias in effective PSF size for different
  calibration methods, when 400 stars that are uniformly sampled from
  the Pickles SED library \citeind{pickles} are used for the training
  step. The dashed-dotted blue line shows the results without further
  calibration, showing a strong redshift dependence. The dotted red
  line indicates the results when the bias is adjusted based on the
  \bi{r}-\bi{i} colour. The dashed green line shows that the bias can
  be reduced significantly when an redshift-dependent offset in the
  \bi{r} magnitude is applied. The solid black line uses a stellar SED
  distribution from fitting to KiDS data (see text). The errors are estimated from 100
  pointings and the horizontal band marks the required accuracy.}
\label{learning_calibrated}
\xef

To examine this in more detail, we train on a star catalog using the
\riz\ bands (see \S\ref{subsec:star_train}) and compute the
residual bias in the effective PSF size as a function of redshift. The
results are presented in Fig.~\ref{learning_calibrated} by the blue
dashed line, and show a clear redshift dependence. The bias is higher
at both low and high redshifts, but these redshift ranges contain fewer
galaxies. When averaging over redshift the biases largely cancel,
leading to a low average value for the full sample.  Such a redshift
dependent bias is problematic as it may mimic an interesting
cosmological signal, and our results demonstrate that it is not
sufficient to specify a requirement for the full sample.

We explored various approaches to model the redshift-dependent bias,
but were unable to do so directly. Instead we opted for a hybrid
approach using a simulated galaxy catalog as a `calibration'
sample. We train on the stars as before, but use the simulated galaxy
catalog to adjust the method to create an unbiased estimate of the
effective PSF size.  Although such an implementation is no longer
fully self-calibrating, training on the observed stars may still
valuable, if it can reduce the sensitivity to errors in the photometric
calibration and the wavelength dependence of the PSF model, as well as
the adopted library of galaxy templates (see \S\ref{calibration} for
more details). On the other hand, the performance may still be limited
by the fidelity of the calibration sample that is used.

One possibility is to adjust the bias in effective PSF size by
accounting for the difference in size as a function of $r-i$ colour
between stars and galaxies, as is indicated in
Fig.~\ref{stars_failure}. The dotted red line in
Fig.~\ref{learning_calibrated} demonstrates, however, that this does
not alleviate the strong redshift dependence. Instead, as is explained
in more detail in Appendix~\ref{app_rshift}, we found that it is
possible to reduce the bias effectively by introducing an \bi{r}-band
offset that depends on redshift, without increasing the scatter
significantly. However, in practice photometric redshifts are used,
and we examine the consequences of this in the next subsection.

The adopted redshift-resolution for this correction allows us to adjust the 
importance of the galaxy templates on the results: if we assume no redshift
dependence the method reverts back to training on stars, whereas a very fine
redshift sampling is identical to training on galaxy templates. As discussed
in Appendix~\ref{app_rshift} we adopted a redshift sampling of $\Delta z = 0.18$, which
appeared to be a reasonable compromise between the two extremes.
In \S\ref{sec:cal_errors} we compare the performance of the hybrid approach
to the training on galaxy templates in the presence of calibration errors in the 
photometric data.

The distribution of stellar SEDs is expected to vary, in particular as a function of Galactic  coordinates.
To explore the impact of such variations we determine the SED distribution by fitting the
Pickles library to the \bi{g}, \bi{r}, \bi{i}, \bi{z} for
stars ($i_{AB}< 21$, $\text{CLASS\_GAL} > 0.8$) in KiDS DR2. For
each {\it Euclid} pointing we simulate 400 stars generated
using the stellar distribution in a KiDS pointing limited to
the 15 most frequent templates. We find that the bias is essentially
unchanged compared to the case of training on a uniform distribution of templates.

\subsection{Correlations between photometric redshift and effective PSF size}
\label{subsec:pzcorr}

\xbf
\xfigure{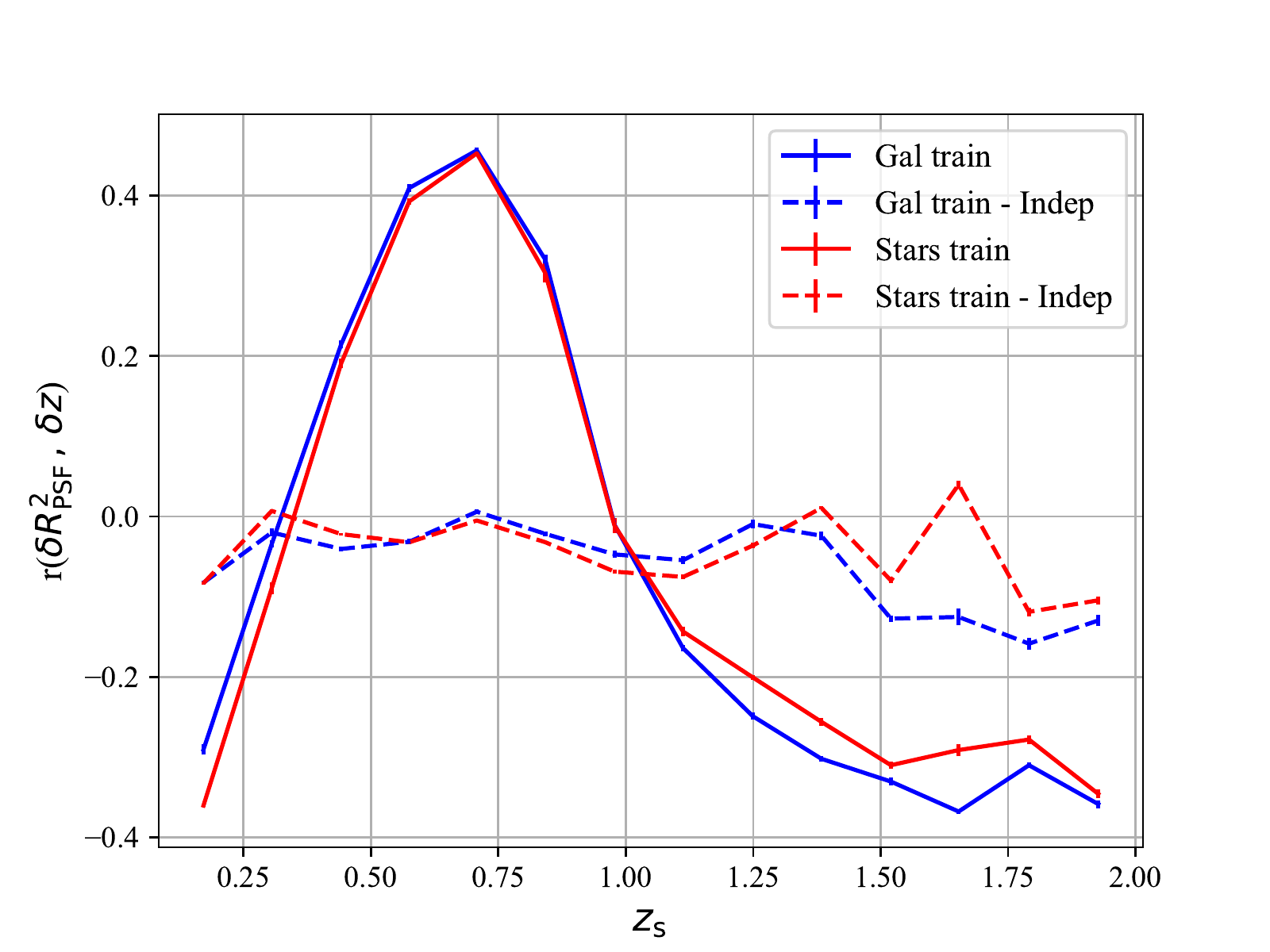}
\xfigure{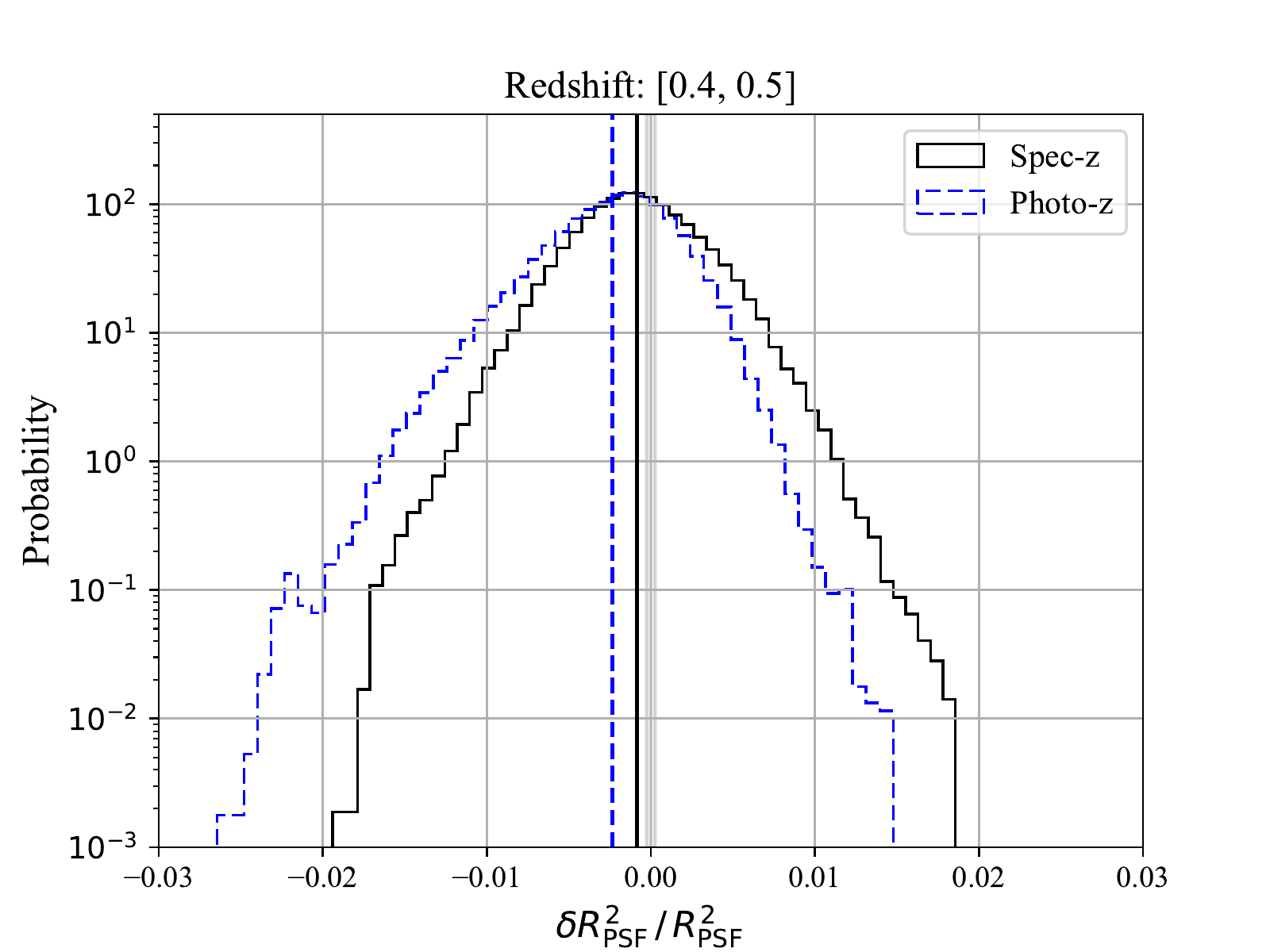}
\caption{\emph{Top:} Pearson correlation coefficient between the
  residual bias in the effective PSF size,
  $\delta R_{\rm PSF}^2/R_{\rm PSF}^2$ and photo-$z$ error, $\delta z$,
  as a function of true redshift. The effective PSF size was estimated
  using the \riz\ bands only, whereas the photometric redshifts use
  measurements from all available filters. The blue lines show the
  results when galaxy mocks are used for the training, whereas the red
  lines show the results when stars are used (but no additional
  calibration applied). The solid lines indicate the results for the
  realistic case when the same \riz\ data are used to measure both PSF
  size and photometric redshift (i.e., the noise is in common),
  whereas the dashed lines are for independent realisations of the
  data. The error bars are estimated from simulating 100 pointings.
  \emph{Bottom:} Histogram of $\delta R_{\rm PSF}^2/R_{\rm PSF}^2$ for
  the redshift range [0.4, 0.5] when selecting either on spec-z
  (solid black) or photo-$z$ (dashed blue). The training uses stars in
  the \riz\ bands and the histograms comprise 100 independent
  pointings.}
\label{pz_corr}
\xef

The results presented in Fig.~\ref{learning_calibrated} show that
the biases as a function of {\it true} redshift can be reduced to the
required level. However, in practice, tomographic bins are based on
the photometric redshifts. The large uncertainties and potential
outliers may affect the performance of the approach outlined
above. Moreover, as the photometric redshift estimate and the
determination of the effective PSF make use of the same data (at least
in part), correlations may be introduced. We explore these more
practical complications here.

To quantify the correlation between the relative error in the inferred
effective PSF size $\delta R_{\rm PSF}^2/R_{\rm PSF}^2$ and the error
in the best fit photometric redshift
$\delta z\equiv (z_{\rm b}-z_{\rm true})/(1+z_{\rm true})$ we define
the Pearson correlation coefficient 

\newcommand{\var}{\text{Var}} 
\be
r(\delta R_{\rm PSF}^2, \delta z) \equiv \frac{\left< \delta R_{\rm PSF}^2, \delta
    z\right>} {\sqrt{\var(\delta R_{\rm PSF}^2) \var(\delta z)}}
\label{rpz_coeff}
\ee

\noindent
where $\var(.)$ is the variance. The solid lines in Fig.~\ref{pz_corr} show the
correlation coefficient between the error in the effective PSF size
and $\delta z$ as a function of true redshift when training on
galaxies or stars, respectively. In both cases we find a significant
positive correlation between the errors between $0.4<z<1$ and an
anti-correlation at high redshifts. In contrast, the dashed lines show
the results when the effective PSF size and photometric redshift
determinations are obtained using independent data sets, i.e.
independent noise realisations.  In this case the correlation
coefficient is close to zero for both training cases.

\xbf
\xfigure{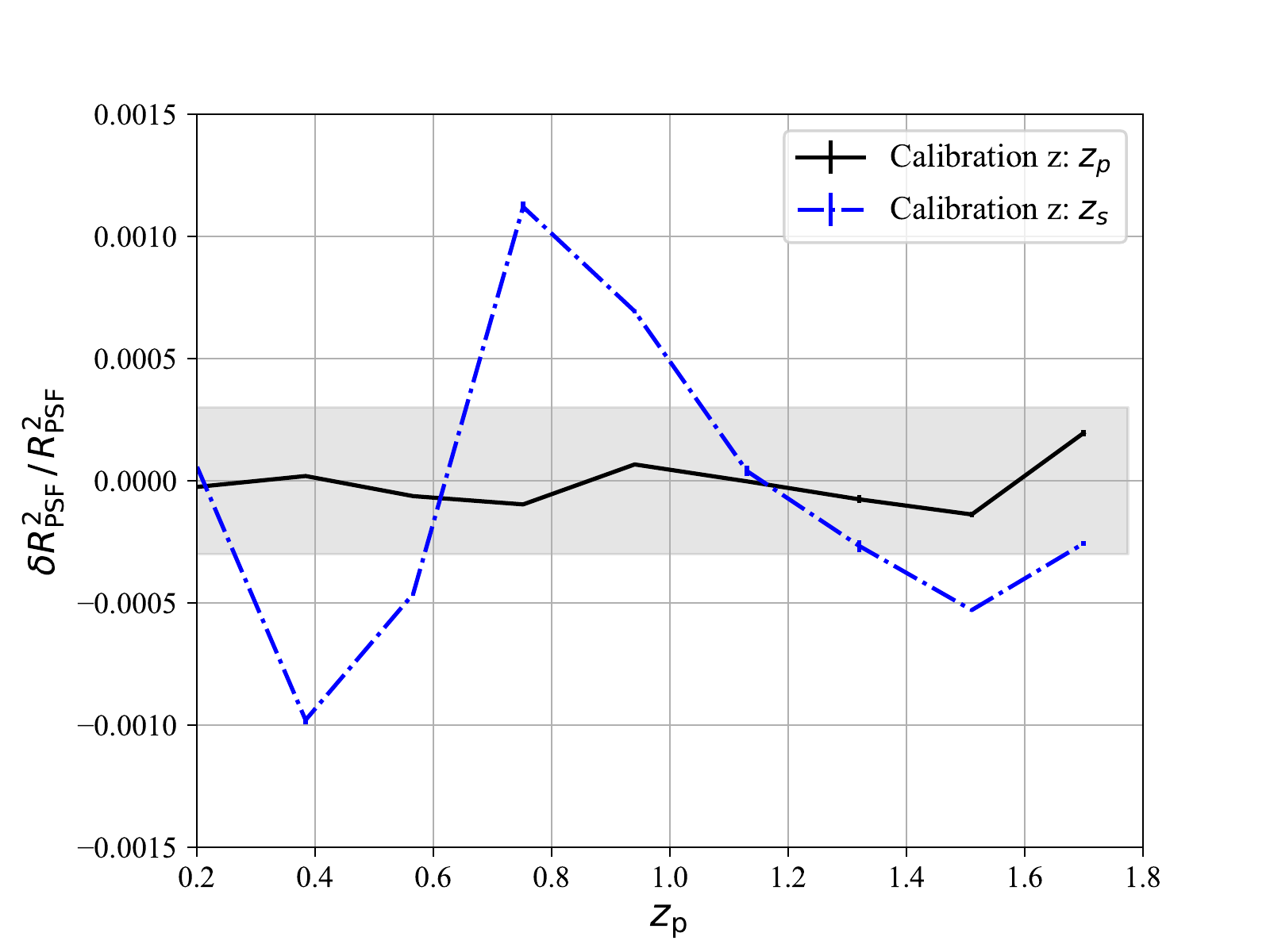}
\caption{$\delta R_{\rm PSF}^2/R_{\rm PSF}^2$ as a function of
  photometric redshift, using either the photometric redshift or the
  true redshift in the calibration catalog. The training uses
  observations of stars in the \riz\ bands and the uncertainties are
  estimated from 100 pointings.}
\label{pz_calib}
\xef

These results demonstrate that the (re-)use of photometric data for
different measurements may lead to redshift dependent correlations
between residuals. To examine this in more detail we consider a single
redshift bin with $0.4<z<0.5$, where the selection can be done either
based on spectroscopic or photometric redshift. The resulting
distributions of the relative bias in effective PSF size are presented
in the bottom panel of Fig.~\ref{pz_corr}.  When the galaxies are
selected by spectroscopic redshift the distribution is symmetric, and
the mean bias meets our requirement.  However, selecting galaxies
based on their photometric redshift leads to a skewed distribution
with a significant bias in the mean. This is the result of the
correlation between $\delta z$ and
$\delta R_{\rm PSF}^2/R_{\rm PSF}^2$: a selection in photometric
redshift leads to a selection in effective PSF size. Hence an unbiased
estimate for tomographic bins needs to account for this correlation.

This is achieved naturally when correcting the bias in effective PSF
size using the simulated galaxy catalog: the redshift dependence of
the applied offset in $r$ magnitude can be determined by splitting the
galaxies as a function of photometric redshift. As shown by the solid
black line in Fig.~\ref{pz_calib}, this removes the impact of the
correlation between the $\delta z$ and
$\delta R_{\rm PSF}^2/R_{\rm PSF}^2$, since this is also included in
the calibration sample.  In contrast, when the calibration sample is
instead split based on the true redshift, $z_s$, the bias exceeds
requirements (blue dashed line). These results indicate that the
calibration step can reduce the bias when using photometric redshifts,
provided that the simulations are sufficiently accurate. We note, however,
that the training needs be done {\it after} the tomographic bins have been
defined.

\section{Calibration}
\label{calibration}
\subsection{Impact of calibration errors.}
\label{sec:cal_errors}

So far we have assumed that the flux measurements in the various bands
used to infer the effective PSF size are perfectly calibrated and that
the wavelength dependence of the PSF is known. In practice the
zeropoints in the \riz\ bands will vary across the survey, although we
note that (some) modern surveys can achieve impressive homogeneity
\citep[e.g.][]{Finkbeiner16}. Moreover, the wavelength dependence of
the PSF is expected to be well-known, but it will vary with time due
to variations in the optical system. 

A machine-learning algorithm trained on the observed stars establishes the 
mapping  between the observed colours and PSF size, rendering it insensitive 
to zero-point offsets  and errors in the PSF model. Interestingly, problems
with the photometry and PSF model could be identified by examining the
observed sizes to the ones expected based on their SED. For instance, the colour-$R^2_{\rm PSF}$
relation is shifted by a magnitude offset and can thus be used to detect 
biases in the ground based photometry. Deviations from the expected wavelength dependence
of the PSF can be directly tested by comparing the observed PSF to the
wavelength dependent PSF model, $R^2_{\rm PSF}(\lambda)$, convolved with stellar spectra.

Unfortunately we  found in the previous section that  it was not possible 
to estimate  the effective PSF using the stars alone, because it resulted 
in strong redshift dependent biases.  We therefore need to quantify the impact of 
calibration errors on the estimate of the effective PSF size. We compare the
sensitivity of the two machine-learning implementations discussed in the previous 
section. Although both approaches ultimately rely on simulated galaxies,
the hybrid method may retain some of the advantages of self-calibration,
and thus the sensitivity to calibration errors may still be reduced. In this subsection 
we use the \riz\ observations of stars to train the algorithm, although we note that the results 
are qualitatively the same when only $r,i$ data are used, albeit with a somewhat larger 
scatter in the latter case.

\xbf
\xfigure{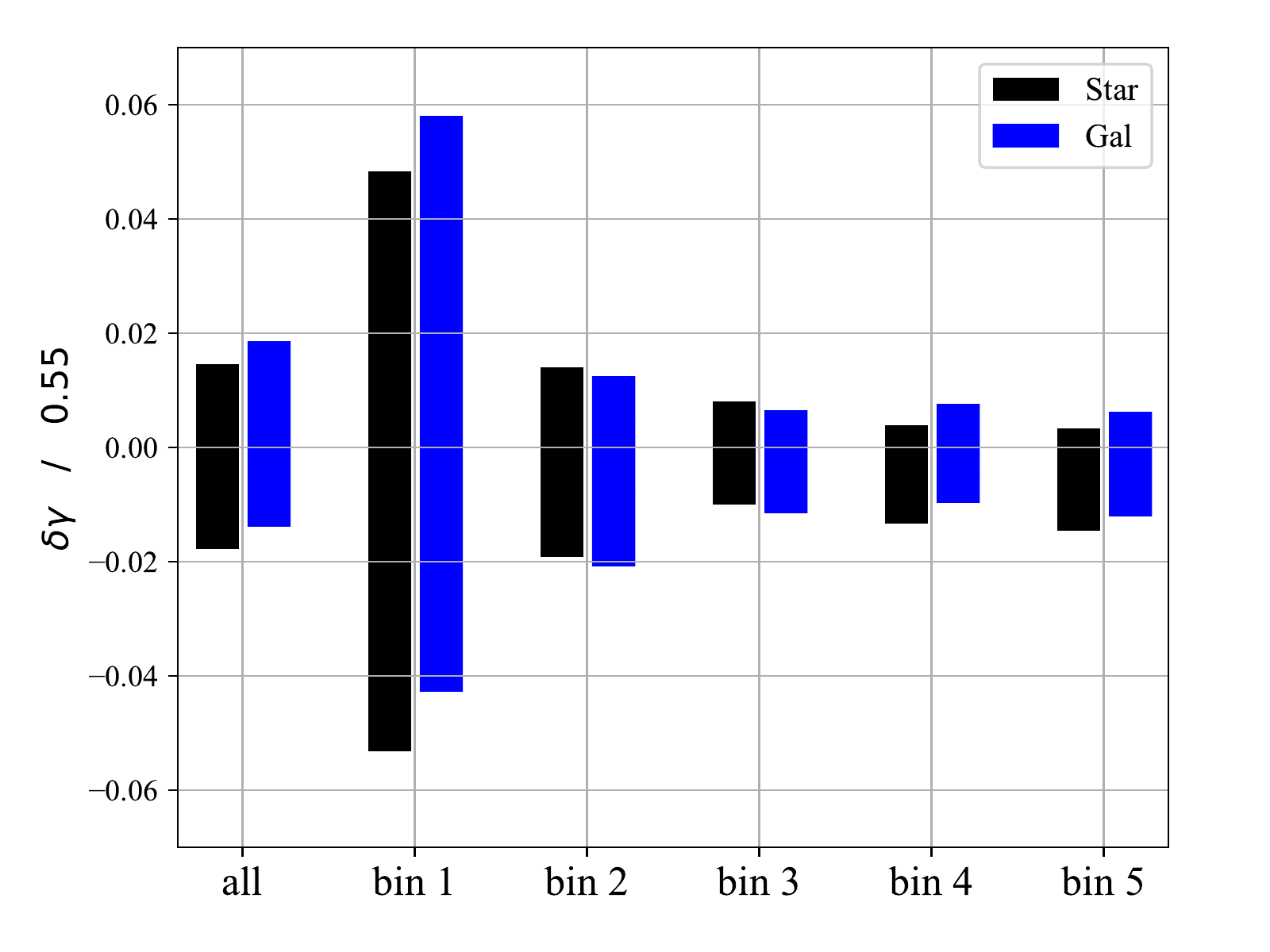}
\caption{The allowed range in offset in the power law slope of
the PSF size as a function of wavelength, $\gamma$, such that the bias 
in effective PSF size does not exceed the {\it Euclid} requirement. The
black bars indicate the results for the hybrid approach, and the blue bars
when we train using galaxy SEDs. To obtain
these results we applied the  offset to the simulated galaxy and star data, while 
the galaxy training and calibration sample use the nominal value of $\gamma = 0.55$.
This corresponds  to an unknown $\gamma$ shift in the data. 
On the y-axis we show the requirement for the full sample (all) and
  when splitting in redshift bins. The five redshift bins are:
  [0.10, 0.44), [0.44, 0.78), [0.78, 1.12), [1.12, 1.46)
  and [1.46, 1.80), in increasing order.}
\label{cal_gammaoffset}
\xef

We first examine the sensitivity to the wavelength of the PSF, which we 
assume to be a power law $R_{\rm PSF}^2\propto \lambda^\gamma$, 
where $\gamma=0.55$ is the nominal value used so far. We assume
that an error in the PSF model is captured by a change in the value of
$\gamma$. For the full sample and different redshift bins, we
find that the relative bias in effective PSF is a linear function of $\delta\gamma$,
the change in the power law slope. We use these relations to determine the
maximum change $\delta\gamma$ that can be tolerated such that the bias in 
effective PSF size is smaller than the {\it Euclid} requirement. The results are shown in 
Fig.~\ref{cal_gammaoffset} for both training approaches. Note that this
shift is applied to the simulated observations, but that the calibration sample
is not changed.

We find that the lowest redshift bin is remarkably insensitive to changes
in the PSF model, and that the  requirements are rather similar for the two 
training approaches. We do note that the galaxy-only case for the fourth
redshift bin represents a challenge: as the sign of $\delta\gamma$ is in
principle unknown, we obtain $|\delta\gamma/0.55|<2.4\times 10^{-3}$.
When we consider the hybrid method that trains on stars
first, we find that $|\delta\gamma/0.55|<5\times 10^{-3}$ is needed to meet the requirement on 
the relative bias in effective PSF size. Although this may appear challenging, 
such a deviation represents noticeable deviation in the optical model of the PSF.
Moreover, the {\it Euclid} PSF is expected to be very stable in time, and thus
changes in $\gamma$ can be easily monitored.

\xbf
\xfigure{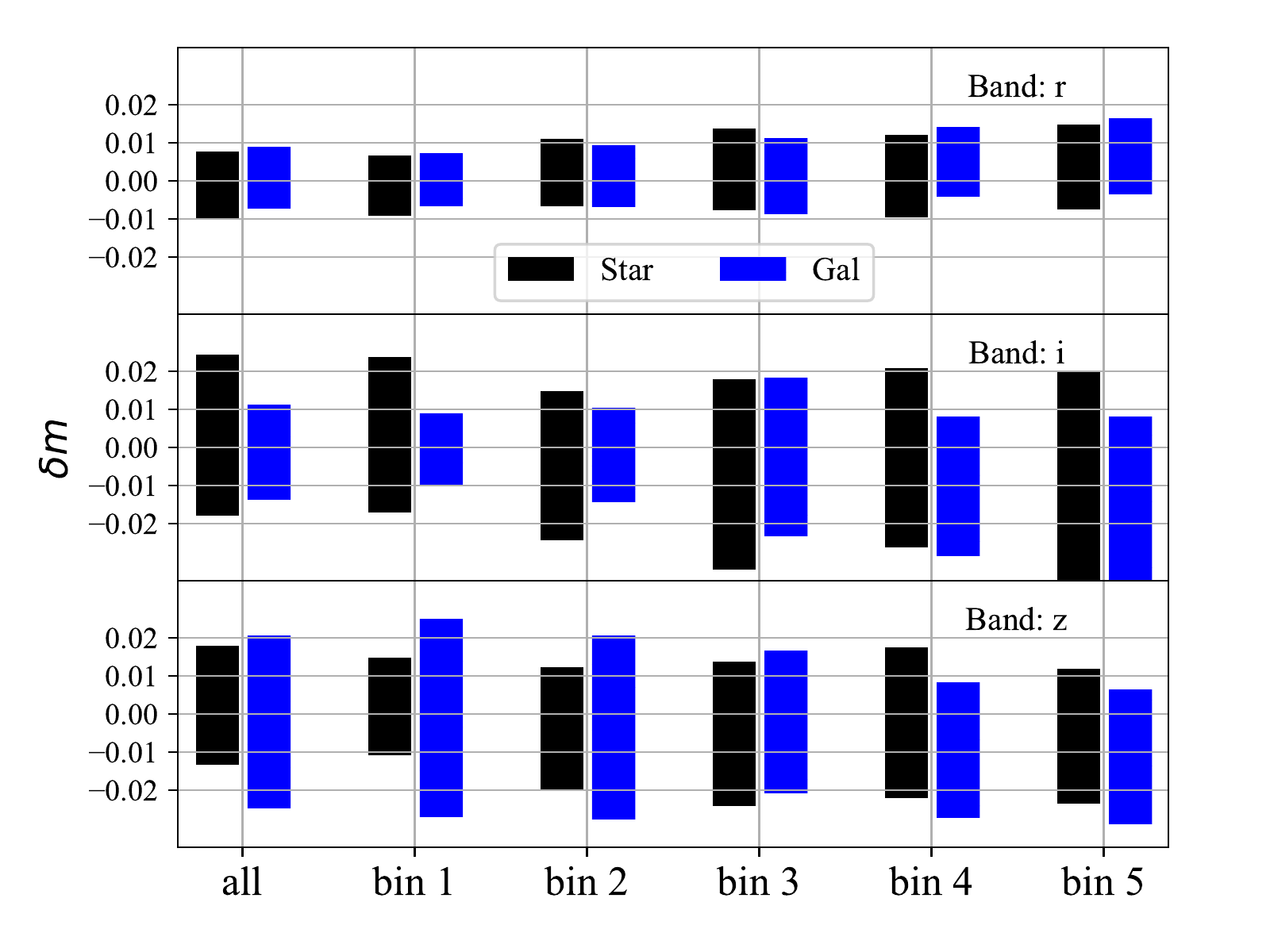}
\caption{The allowed range in magnitude offsets that can be tolerated
for {\it Euclid}. The black bars indicate the results for the hybrid method,
and the blue bars when training on galaxies only. In both cases only
  \riz\ data were used to train on. In the top, middle and bottom panels, 
  the offsets are applied to the \bi{r},\bi{i} and \bi{z}-band independently. On
  the y-axis we show the requirement for the full sample (all) and
  when splitting in redshift bins. The five redshift bins are:
  [0.10, 0.44), [0.44, 0.78), [0.78, 1.12), [1.12, 1.46)
  and [1.46, 1.80), in increasing order.}
\label{cal_magoffset}
\xef

We now proceed to examine the sensitivity to errors in the calibration
of the ground-based observations, which may occur because of varying
observing conditions. Repeated observations of the same field
allow for exquisite homogeneity \citep{Finkbeiner16}. In the near
future, however, {\it Gaia} \citeind{gaia} spectrophotometric measurements 
should provide an excellent reference across the full sky in this wavelength range.
Nonetheless it is important to evaluate what errors in photometric
zeropoint can be tolerated. 

To examine this we apply an offset to the simulated observations in each of 
the optical bands used  and calculate the relative changes 
in the effective PSF size. To do so, we change the zeropoint in one
filter while keeping the other measurements unchanged, and 
determine the maximum shifts in zeropoint that are still within
requirements and show the results in Figure~\ref{cal_magoffset} for the full 
sample of galaxies and the five tomographic bins in true redshift.
The results show that the effective PSF is most sensitive to photometric
calibration errors in the $r$ band: for the hybrid method we find
that  $|\delta m|<0.005$  in the $r$-band, whereas $|\delta m|<0.01$ appears 
adequate for the other filters.  Such a level of photometric homogeneity seems quite
achievable. Training on galaxy simulations alone would increase the sensitivity to
magnitude offsets in the \bi{r}-band such that $|\delta m| < 0.003$.
Hence the hybrid method of training on stars, followed by a redshift
dependent adjustment determined using simulated galaxies retains some 
of the self-calibration properties. 

\subsection{Sensitivity to template library}
\label{subsec:detect_cal}

\xbf
\xfigure{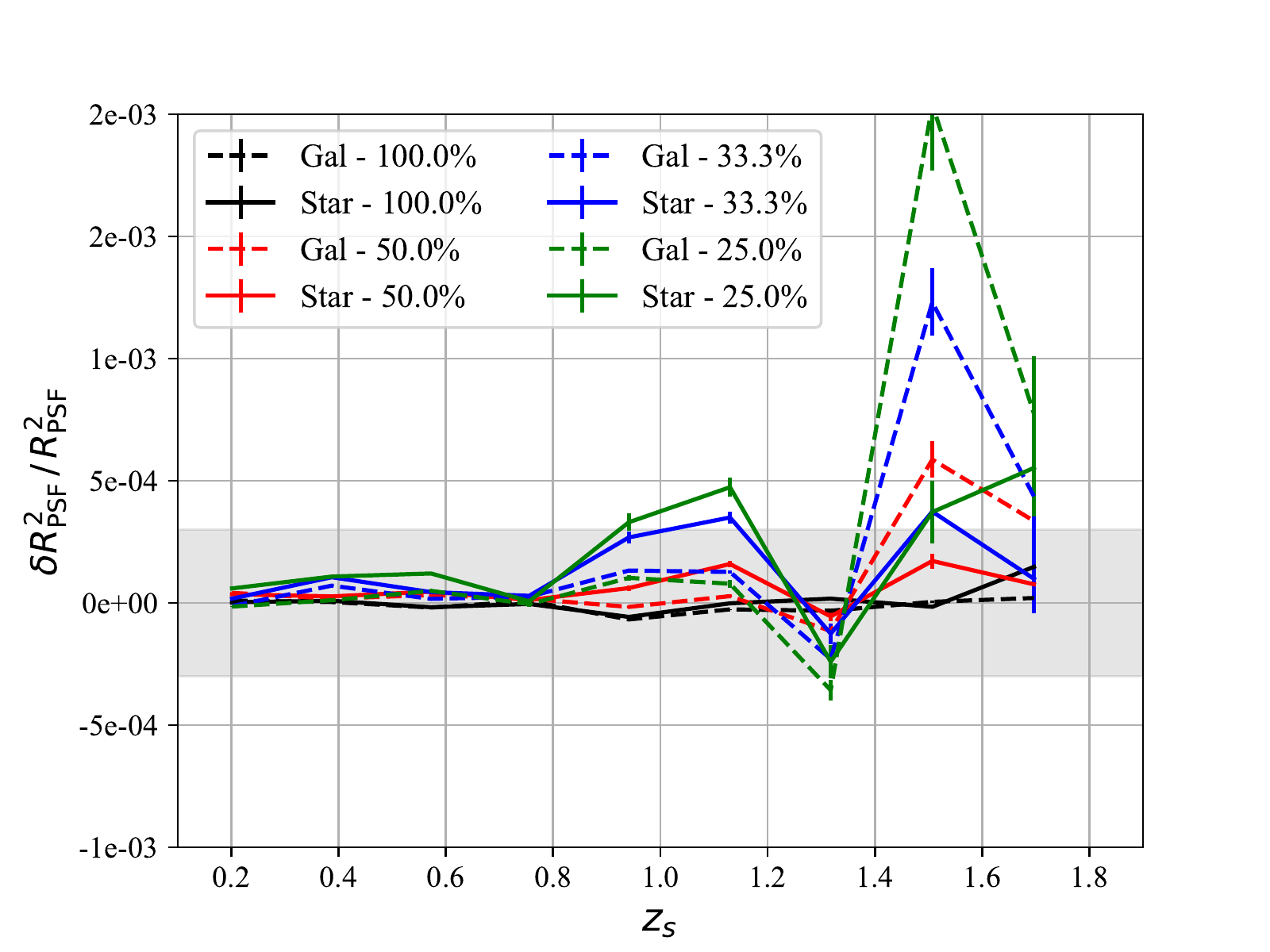}
\xfigure{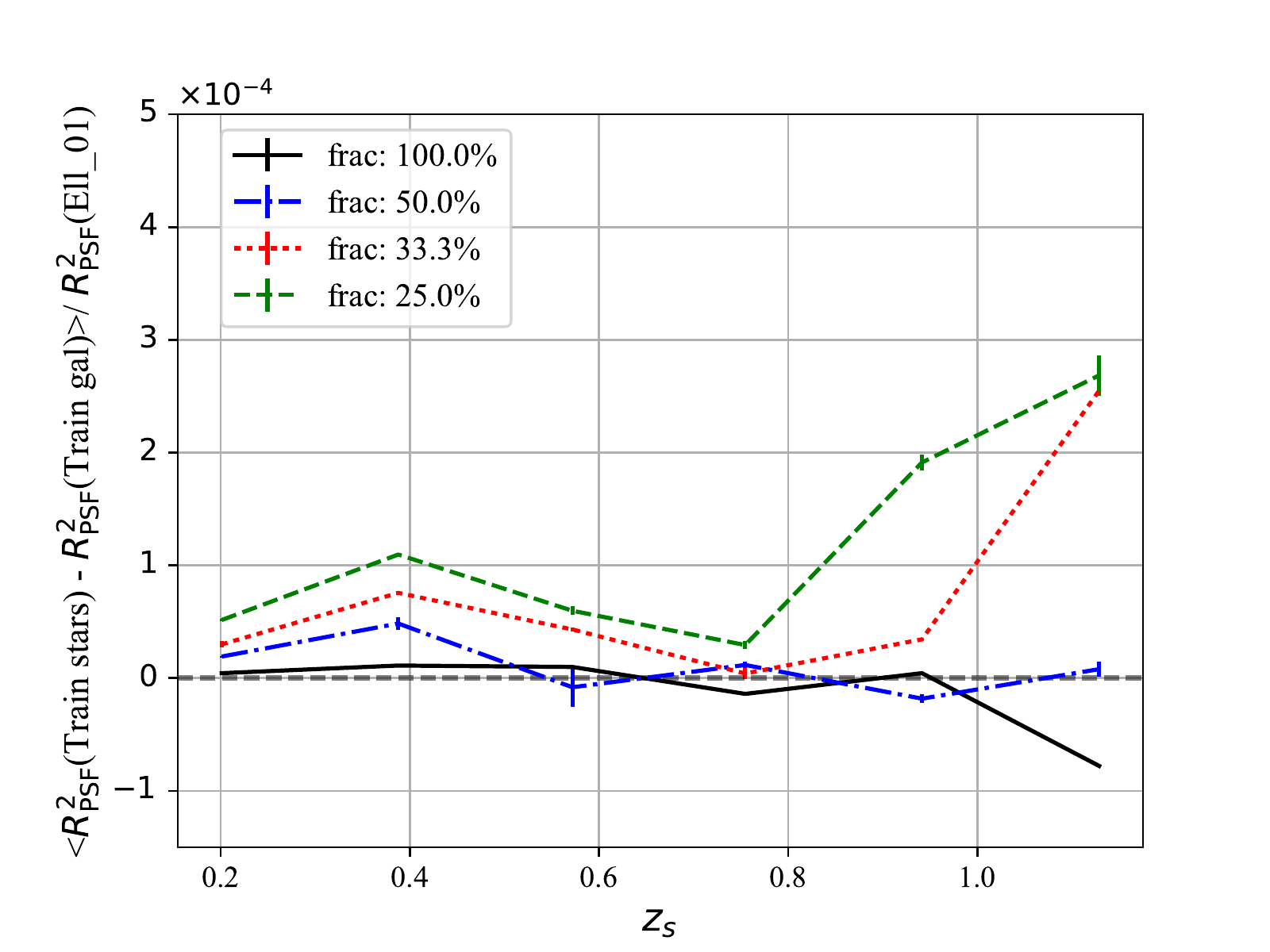}
\caption{\emph{Top:} The relative $R^2_{\rm PSF}$ bias when limiting the
SEDs in the galaxy training and the calibration sample. Solid and dashed
lines show $\delta R^2_{\rm PSF} \, / \, R^2_{\rm PSF}$ when training
on stars and galaxies, respectively. The template removal is done uniformly
over the 66 SEDs and legend shows the fraction remaining. On
the x-axis it the spectroscopic redshift. \emph{Bottom:} The mean difference
between the $R^2_{\rm PSF}$ predicted when training on simulated observed 
stars and simulated galaxies. The errorbars for both panels are estimated
from 100 {\it Euclid} pointings.}
\label{detect_main}
\xef

Given that both machine-learning approaches rely on a library of simulated
galaxy SEDs, a key remaining question is whether the results are sensitive
to limitations of the template library. In \S\ref{subsec:sedfit} we already
saw that the performance of the template fitting code depends on the
number of templates used, especially if the number was too low. We therefore
return to this issue here and examine the impact of an incomplete SED
library on the machine-learning approaches. 

The top panel in Fig.~\ref{detect_main} shows $\delta R^2_{\rm PSF} \, / \, R^2_{\rm PSF}$
when we limit the SED coverage in the galaxy training sample and
the calibration sample used to adjust the predictions when training on stars first. 
The legend shows the fraction of SEDs which remain after uniformly removing 
templates from the simulations. For a complete sample ($100\%$), the 
bias in effective PSF size is small for both approaches. When omitting galaxy templates, 
the hybrid method performs better at high redshift. Overall, both methods
appear fairly robust to an incomplete template library, although this may need
to explored further; especially the impact of emission-line galaxies and dusty
galaxies has not been explored.

As the hybrid method requires the same template library, one may be tempted
to omit this approach. Regardless, we need to somehow ensure that the
template library is adequate. This can be done by comparing the results
from the two methods, as their dependencies on the template library is 
different. This is evident from the top panel in  Fig.~\ref{detect_main}. The bottom 
panel of Fig.~\ref{detect_main} shows the difference in the estimate of the effective PSF size, 
relative to that of an early type galaxy, as a function of redshift. If the template library is highly 
incomplete, the two PSF estimates differ significantly, especially for high redshift galaxies.
This can thus be used as a way to validate the machine-learning approach.

\section{Conclusion}
The convolution of galaxy images by the PSF is the dominant source of
bias for weak gravitational lensing studies, and an accurate estimate
of the PSF is thus essential for a successful measurement.  In this
paper we studied the bias caused by the combination of a wavelength
dependent PSF and limited information about the SED of a galaxy of
which we wish to measure the shape. We quantified the impact on the
performance of {\it Euclid}, for which the impact is exacerbated
because of the combination of a (near-)diffraction limited PSF and the
broad VIS pass-band. We note, however, that the
wavelength dependence of the PSF cannot be ignored for other stage IV
ground-based surveys such as LSST.

Based on the analysis of biases by \cite{massey13}, we restricted the
study to the determination of the effective (or SED-weighted) PSF
size, which is different for each galaxy. Under the assumption that an
accurate model of the PSF as a function of wavelength can be derived,
we explored several approaches to estimate the effective PSF size from
a number of broad-band images.  Given the exquisite precision with
which {\it Euclid} can measure the cosmic shear signal, the corresponding
accuracy with which the PSF properties need to be determined makes
this challenging. Following \cite{cropper} we consider a maximum
relative error of $\euclidreq$, whereas the variation in effective PSF
size with source redshift is two orders of magnitude larger (see
Figure~\ref{r2_zdep}).

We simulated catalogs of stars and galaxies based on the expected depth and 
wavelength coverage of DES and {\it Euclid}. We used these to examine how 
well a standard template fitting photo-$z$ code can predict the effective PSF for a
galaxy. We found that the resulting distribution of predicted PSF sizes is skewed,
leading on biases in the estimate of the effective PSF that exceed the requirements.
Hence, modifications to the template fitting codes are needed if these are to be
used for this purpose. We note that this may be worthwhile, because the
effective PSF can be computed using a multi-wavelength model of the spatially
varying PSF and the SED from a template fitting code. 

To quantify the expected performance for different scenarios of multi-wavelength
data, we used machine-learning methods instead. We used a NuSVR algorithm to
train on simulated galaxy catalogs and found  that it is possible to reduce the skewness in the predicted effective PSF sizes, 
and thus also the bias in the mean value. We considered various filter configurations and
found  that good results can be obtained when using the $\bi{r}-\bi{i}$ colour only.
A potential complication is the fact that part of the photometric data
are used to estimate both the effective PSF size and the photometric
redshift of a source.  This leads to a correlation between the error
in PSF size and photometric redshift: galaxies with a large error in
PSF size are also more likely to migrate redshift bins when binned in
photometric redshift, causing a selection bias.  Fortunately, we found that when the training is
also done using photometric redshifts, the correlation is naturally
accounted for.

We also examined whether it is possible to predict the effective PSF size using observations of stars in the
data. Such an approach would be immune to photometric calibration errors. To test this, we trained and 
applied a NuSVR algorithm on simulated data. Interestingly, training on \riz\ observations of stars resulted in sufficiently small residuals for the full sample, but the bias varied with redshift significantly. To account for this, we introduced a correction  based on a calibration sample of simulated galaxies.  The adopted redshift-resolution for this correction 
allows us to adjust the importance of the galaxy templates on the results: if we assume no redshift 
dependence the method reverts back to training on stars, whereas a very fine redshift sampling 
is identical to training on galaxy templates.

In the case of perfectly calibrated data the two machine-learning implementations
are similar, but in the presence of calibration errors the performances may differ. 
We examined the sensitivity to errors in the PSF model and the photometric calibration,
and found a slightly better performance for the hybrid method. In this case the power-law 
slope of the wavelength dependence of the PSF size needs to be known to better than
$|\delta\gamma|<5\times 10^{-3}$ which is quite achievable. Moreover,
the estimated effective PSF size is most sensitive to the zeropoint errors in
the $r$-band, which needs to be accurate to $|\delta m|<0.005$, whereas
$|\delta m|<0.01$ is sufficient for the $i$ and $z$ bands. Such levels
of photometric homogeneity have already been achieved, and we expect
the situation to improve further thanks to spectrophotometric
observations with {\it Gaia}.

As both implementations rely on a simulated template library, we explored whether limitations 
of the template library may pose a problem. The hybrid method
is less sensitive to an incomplete set of SEDs, but given our knowledge of 
galaxy SEDs this does not seem to result in a major bias. However, further work
may be needed to examine the impact of emission line galaxies, which we did
not consider here. Although we note this caveat, we conclude that it is possible to 
estimate the effective  PSF to the required level of accuracy for {\it Euclid} with 
the anticipated photometric data.

\section*{Acknowledgments}
M.E. and H.H. acknowledge support from the European Research Council under FP7
grant number 279396. The authors thank 
Jerome Amiaux,
Mark Cropper, Jean-Charles Cuillandre, Thomas Kitching,
Yannick Mellier, Jason Rhodes and Gijs Verdoes Kleijn for discussions and comments.
In this project we used the
\textsc{scikit-learn} \citeind{scikitlearn},
\textsc{pandas} \citeind{pandas},
\textsc{scipy} \citeind{scipy},
\textsc{matplotlib} \citeind{matplotlib},
and \textsc{ipython} \citeind{ipython} software packages.
This research has made use of NASA's Astrophysics Data System.

\appendix
\section{Redshift resolution}
\label{app_dzresol}
\begin{figure}
\xfigure{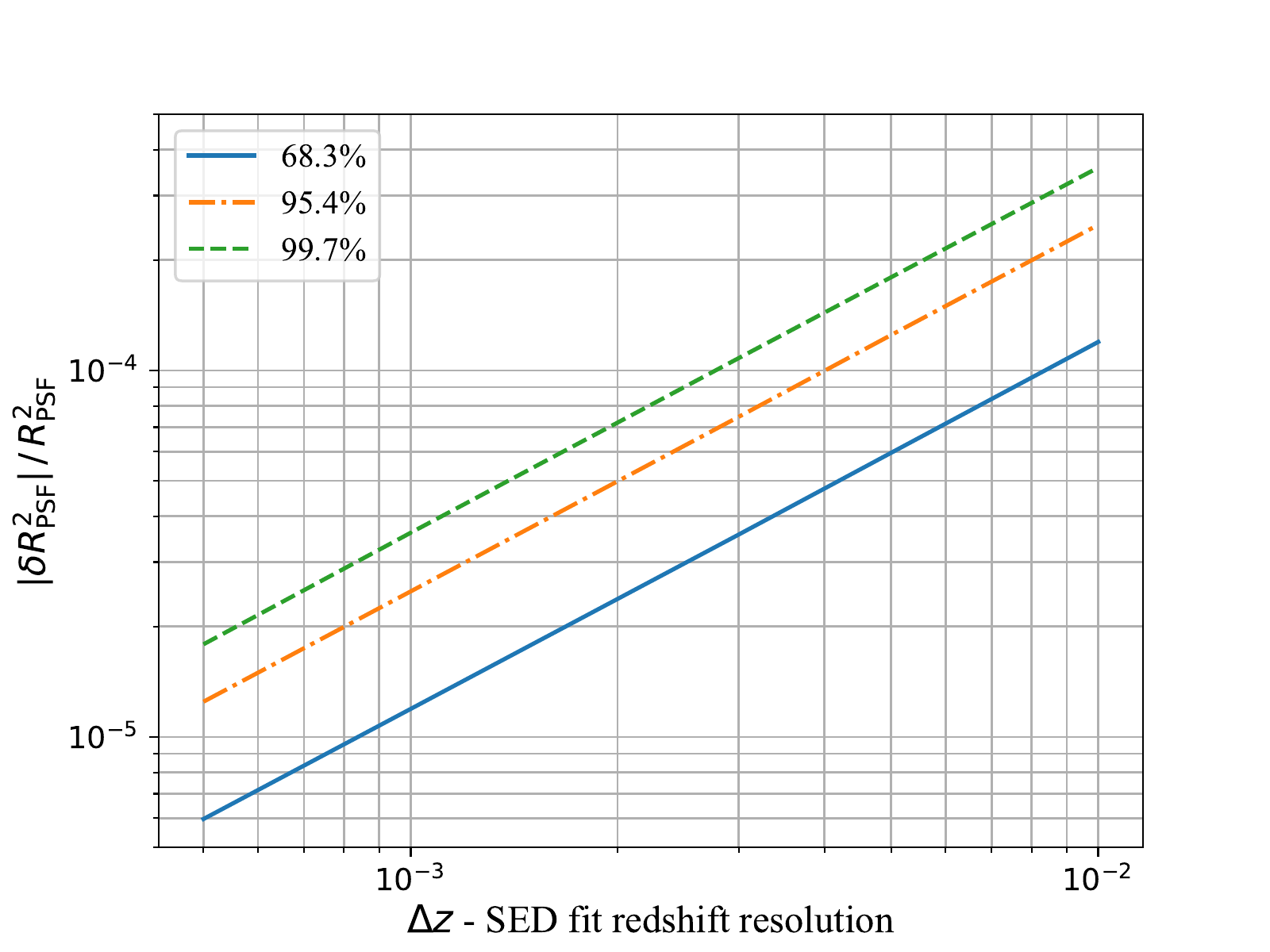}
\caption{The relative error in the effective PSF size $R_{\rm PSF}^2$ as a
  function of the redshift resolution used in the SED template fitting
  method. $R_{\rm PSF}^2$ is estimated using the \riz\ bands with no
  magnitude noise and without photo-z priors. The three lines show the
  $\setlength{\medmuskip}{0mu} 1-\sigma, 2-\sigma$ and
  $\setlength{\medmuskip}{0mu} 3-\sigma$ limits of the
  $\delta R_{\rm PSF}^2$ distribution.}
\label{r2pz_resol}
\end{figure}

Template fitting algorithms \citep[e.g. BPZ;][]{benitez2000} compare
observations with a model on a two-dimensional grid of redshift and
galaxy SED. The number of SEDs depends on the number of templates used
and the number of interpolations between consecutive templates (we use
2). Given the rather poor precision in redshift
($\sigma_z\sim 0.03-0.05$) that can be achieved using broad-band data,
the redshift sampling is typically $\Delta z \approx 0.01$ in order to
minimize runtime. Increasing the redshift resolution is
straightforward as it only requires modifying a configuration
parameter, but the runtime increases roughly linearly with the number
of evaluations in redshift. Given the sensitivity to photometric
redshift errors (see \S\ref{subsec:pzgauss}) we check here if the
default setting is sufficient to estimate the effective PSF size.

To do so, we determine $\delta R^2_{\rm PSF}/R^2_{\rm PSF}$ as a
function of the redshift resolution $\Delta z$. To isolate the effect
of the redshift resolution we consider an idealised situation with no
measurement errors and fit the simulated \riz\ data without photo-z
priors. For $\Delta z = 0.01$ the resulting average relative bias is
below $10^{-6}$ (not shown), which is two orders of magnitude below requirements.
Instead we show in Fig.~\ref{r2pz_resol} the
$\setlength{\medmuskip}{0mu}1-\sigma, 2-\sigma, 3-\sigma$ limits of
the $\delta R_{\rm PSF}^2$ distribution, which also captures the
tails of the distribution of $\delta R_{\rm PSF}^2$ values. Fitting with a resolution
$\Delta z = 0.01$ is thus sufficient to avoid introducing an
additional bias, which we also verified with noisy simulations.

\section{$R^2$ bias calibration technique}
\label{app_rshift}
\xbf
\setlength\floatsep{1.25\baselineskip plus 3pt minus 2pt}
\xfigure{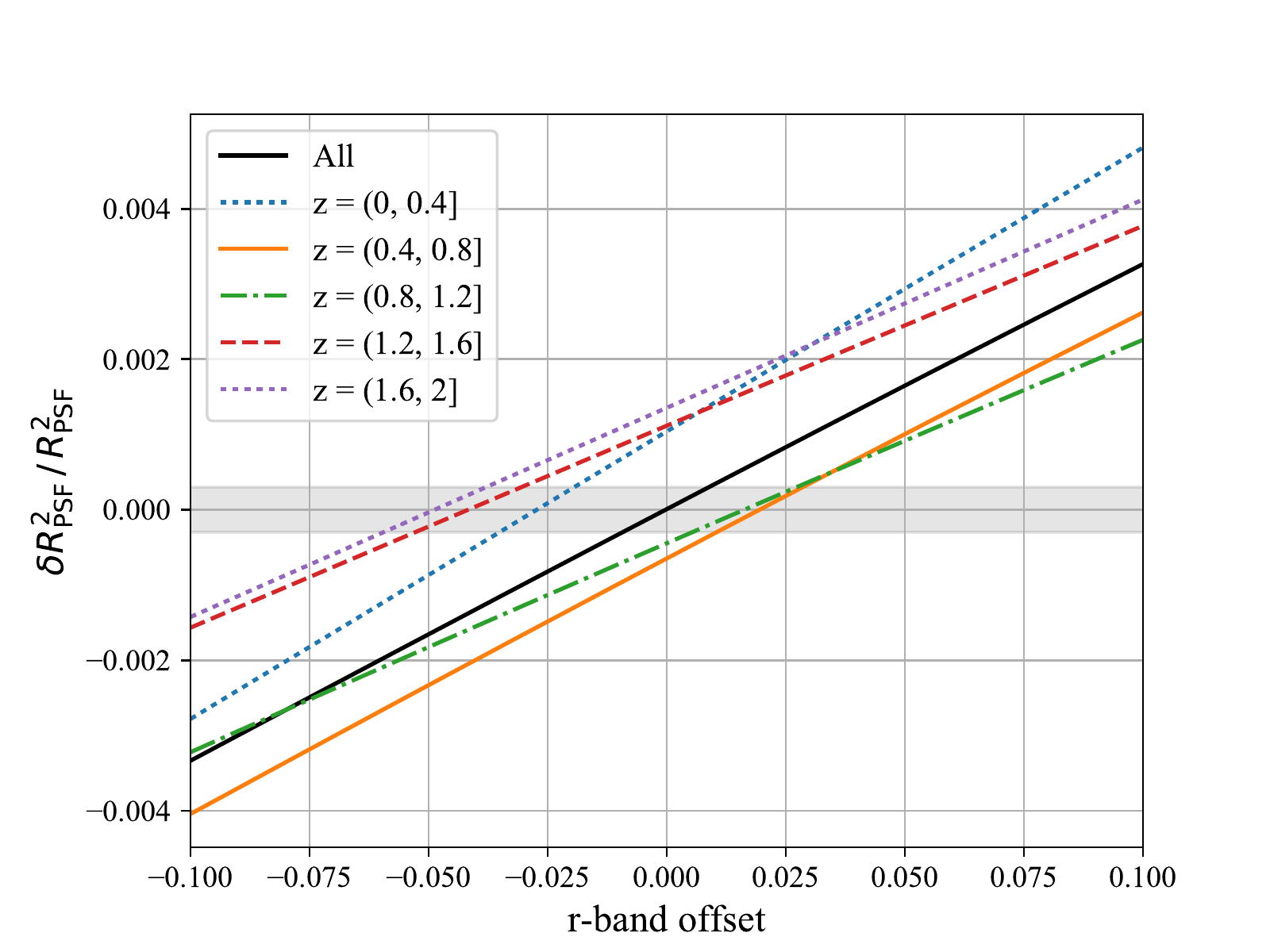}
\xfigure{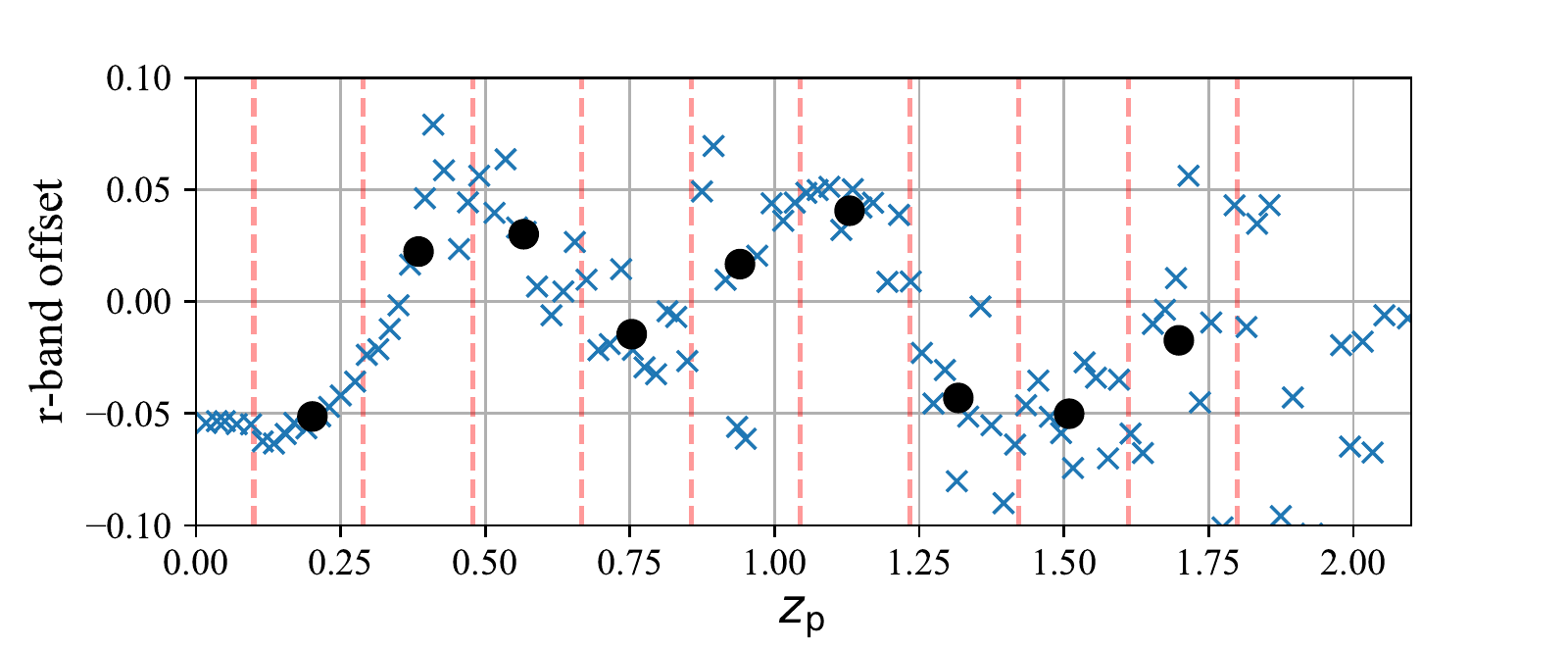}
\caption{\emph{Top:} 
$\delta R^2_{\rm PSF}/R^2_{\rm PSF}$ in the calibration sample as a function
of the applied \bi{r}-band offset. The training uses the
\bi{r}-\bi{i} and \bi{i}-\bi{z} colours from stars and the offset is
included in the \bi{r}-\bi{i} colour. One line shows the full sample
(All), while the rest split in photometric redshift bins. A shaded
horizontal band marks the {\it Euclid} requirement.  \emph{Bottom:} The
optimal \bi{r}-band offset as a function of redshift. The circles is the r-band
offset in the bins ($\Delta z=0.18)$ marked with vertical lines, while the crosses
use a resoltion of $\Delta z=0.02$.}
\label{rband_shift}
\xef

The \textsc{scikit-learn} library includes many algorithms for
regression.  Several other algorithms (tree based and nearest
neighbours) performed well for the full sample of galaxies when
applied to \riz\ observations of stars.  Unfortunately, all resulted in
a strong variation of the bias in effective PSF size with redshift,
resulting in the need for an additional correction. There are multiple
avenues to reduce these residual biases using a calibration sample of
simulated galaxies.  For instance, one could determine the offsets as a
function of color or redshift. However, reliably predicting a small
bias based on noisy input is difficult. We explored various
approaches, but were unable to achieve the required performance.

Interestingly, we found that introducing a small \bi{r}-band magnitude
offset in the algorithm to correct the predicted value of $R^2_{\rm
PSF}$ yielded satisfactory results.  Figure~\ref{rband_shift} shows a
linear dependence of $\delta R^2_{\rm PSF}/R^2_{\rm PSF}$ to an
\bi{r}-band shift for different redshifts when training on stars with
the \riz\ bands. We found good performance when we consider
nine redshift bins; as shown in the bottom panel of Fig~\ref{rband_shift} the resulting \bi{r}-band 
offset varies fairly smoothly with redshift. A finer binning in redshift
(as indicated by the crosses) shows that the actual variation with redshift
is more erratic. This is naturally captured when training on galaxy templates
alone. Hence the redshift sample $\Delta z$ can be considered a parameter
that regulates the importance of the galaxy templates: no binning corresponds
to training on stars, whereas $\Delta z=0$ is equivalent to training on the
galaxy templates. We found that our choice of $\Delta z=0.18$ provided a good
compromise that performs well for the prediction of the effective PSF size of a galaxy.

\section{Performance without \bi{z}-band observations}
\label{app_zband}
\xbf
\xfigure{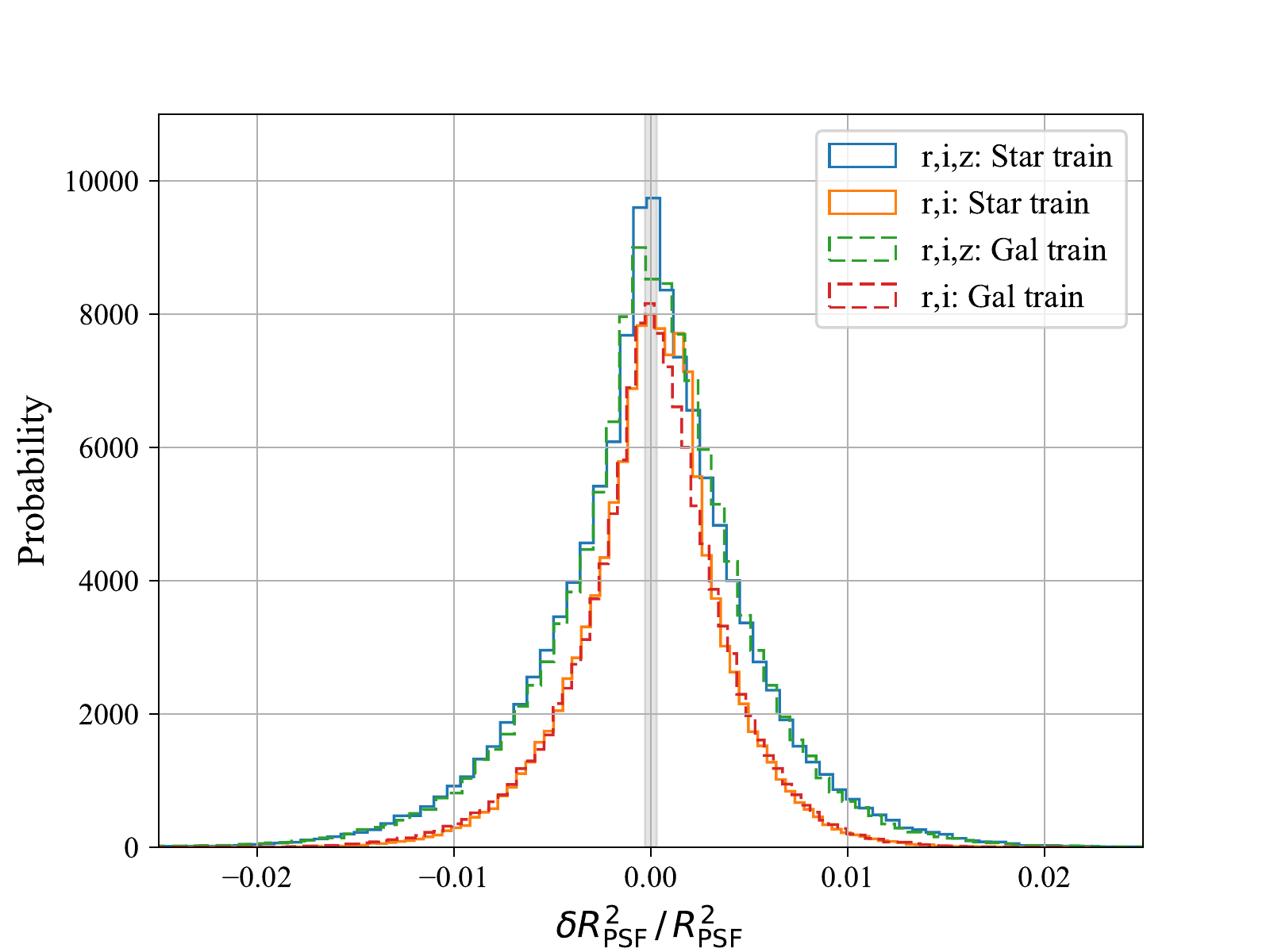}
\caption{Histogram of $\delta R_{\rm PSF}^2/R_{\rm PSF}^2$ values when
training using either \riz\ (blue) or \bi{r} and \bi{i} (red) data.
The solid lines correspond to the case where the algorithm is trained
on stars, whereas the dashed lines show the results when training on
the galaxy simulations. The vertical band marks the {\it Euclid}
requirement for the mean bias.}
\label{riz_ri_hist}
\xef

When training on galaxies, the best performance was obtained when
we used $r$ and $i$ photometry only, which can be understood because 
the VIS passband overlaps only with the blue half of the full $z$-band. 
It is therefore interesting to examine whether $z$-band data can be omitted, 
especially given the relatively large amount of observing time needed to obtain 
these data from the ground.

To compare the performances of the various filter combinations, we
show in Fig.~\ref{riz_ri_hist} the distributions of $\delta R_{\rm
PSF}^2/R^2_{\rm PSF}$. The main impact of omitting the $z$-band is an
increase in the scatter, but this has a negligible impact on the
overall precision of the weak lensing analysis.  We note that the
approach discussed in Appendix~\ref{app_rshift} also works well using
only $r$ and $i$ data (see Fig.~\ref{mag_depth}).

\xbf
\xfigure{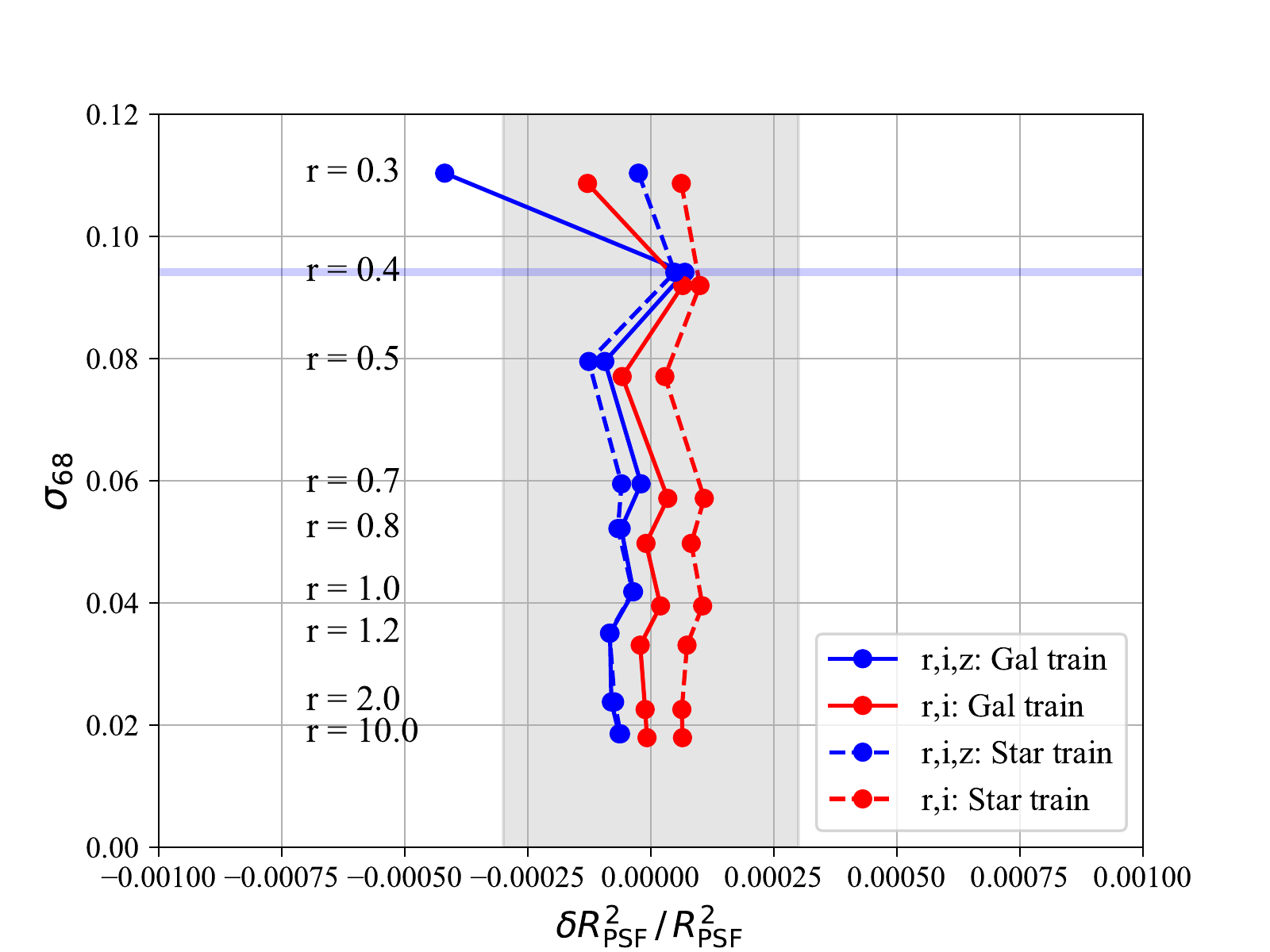}
\caption{The effect of the magnitude depth on the photometric redshift and
the $R_{\rm PSF}^2$ bias. The points indicate the ratio in the
exposure times relative to the fiducial setup (with values indicated).
The blue lines indicate the results when training using the \riz\ data,
whereas the red lines are for the case where the \bi{z}-band is
omitted. The solid lines are for the results when the algorithm is
trained on stars, and the dashed lines are when the simulated galaxies
are used to train. Note that when the \bi{z}-band is omitted, it is
also not used in the photo-$z$ determination. The vertical band marks
the {\it Euclid} requirement bias in effective PSF size.}
\label{mag_depth}
\xef

The depth of the supporting multi-band photometry for {\it Euclid} is
determined by the need for sufficiently precise photometric redshifts,
but also affects our ability to determine the effective PSF size. It
is therefore important to examine the impact of changes in depth (and
filter coverage) on both of these key aspects of the lensing
measurements.

This is captured in Fig.~\ref{mag_depth} which shows $\sigma_{68}$
and $\delta R_{\rm PSF}^2/R_{\rm PSF}^2$ as a function of exposure time, where
$\sigma_{68}$ is the (average two-sided) 68 per cent limit of $(z_p -
z_s)$, which corresponds to $1-\sigma$ for a Gaussian distribution,
but is less sensitive to the outliers than the rms. The magnitude
errors enter in the estimates of the photometric redshift and the
effective PSF size. In the figure `r' indicates the relative change in
exposure time in the \riz\ bands, relative to the nominal values used
throughout the paper. The photo-$z$ estimate does not include the
{\it Euclid} VIS-band. Not surprisingly, longer exposures
significantly decrease the photo-$z$ scatter and vice versa.  We also
show results without $z$-band observations and find only small
differences in the precision of the photometric redshifts and the bias
in effective PSF size.

The most noticeable result is how insensitive $\delta R_{\rm
PSF}^2/R_{\rm PSF}^2$ is to the change in depth. For the fiducial
exposures all lines are well within the requirement. This can be
attributed to the calibration sample, which also simulates the
measurement noise: when changing the exposure time we also change the noise
level in the calibration sample and the calibration therefore helps to
remove the noise bias. We note, however, that this does increase the
noise in the PSF estimate.  Nonetheless it is clear that the required
precision in the determination of photometric redshifts is the main
driver for the depth of the photometric data.

\bibliography{bibpsfw}{}
\bibliographystyle{mn2e}

\end{document}